\documentclass[conference]{IEEEtran}
\usepackage[a4paper,left=3cm,right=2cm,top=2.5cm,bottom=2.5cm]{geometry}
\usepackage{cite}
\usepackage[pdftex]{graphicx}
\usepackage[cmex10]{amsmath}
\usepackage{amssymb}

\usepackage{times}
\usepackage{array}
\usepackage[tight,footnotesize]{subfigure}
\usepackage[font=footnotesize]{subfig}
\usepackage{graphicx}
\usepackage{mathtools}
\usepackage{amsmath}

\usepackage{alltt}
\usepackage{shadethm}
\newshadetheorem{comment}{Comment}
\newtheorem{lemma}{Lemma}
\newtheorem{theorem}{Theorem}

\def\boxx{{\vcenter{\vbox{\hrule height.3pt
          \hbox{\vrule width.3pt height6pt
          \kern6pt\vrule width.3pt}\hrule height.3pt}}\;}}

\def\impos{{\;\vcenter{\hbox{\rule{5mm}{0.2mm}}} \vcenter{\hbox{\rule{1.5mm}{1.5mm}}} \;}}

\def\lrarrow{\leftrightarrow \kern-8pt \rightarrow}

\def\Unspec{A_{\rm ?}}

\def\2{\frac{1}{2}}

\def\beq{\begin{eqnarray}}
\def\eeq{\end{eqnarray}}
\def\2{\frac{1}{2}}

\newtheorem{assumption}{Assumption}

\newtheorem{example}{Example}

\newtheorem{definition}{Definition}

\def\lrarrow{\leftrightarrow \kern-8pt \rightarrow}

\def\frightarrow{\rightarrow \kern-11pt /~~}
\def\reducesto{\simeq \kern -3pt >}
\def\intersection{\cap}
\def\therefore{\stackrel{\rm implies}{\longrightarrow}}

\begin{document}
\newcommand{\strust}[1]{\stackrel{\tau:#1}{\longrightarrow}}
\newcommand{\trust}[1]{\stackrel{#1}{{\rm\bf ~Trusts~}}}
\newcommand{\promise}[1]{\xrightarrow{#1}}
\newcommand{\revpromise}[1]{\xleftarrow{#1} }
\newcommand{\assoc}[1]{{\xrightharpoondown{#1}} }
\newcommand{\rassoc}[1]{{\xleftharpoondown{#1}} }
\newcommand{\imposition}[1]{\stackrel{#1}{\impos}}
\newcommand{\scopepromise}[2]{\xrightarrow[#2]{#1}}
\newcommand{\handshake}[1]{\xleftrightarrow{#1} \kern-8pt \xrightarrow{} }
\newcommand{\cpromise}[1]{\stackrel{#1}{\frightarrow}}
\newcommand{\policy}{\stackrel{P}{\equiv}}
\newcommand{\field}[1]{\mathbf{#1}}
\newcommand{\bundle}[1]{\stackrel{#1}{\Longrightarrow}}

\title{Locality, Statefulness, and Causality\\in Distributed Information Systems\\~\Large Concerning the Scale Dependence Of System Promises}
\author{Mark Burgess\\~\\Aljabr Inc.\\ChiTek-i AS\\~}
\maketitle
\IEEEpeerreviewmaketitle
\thispagestyle{plain}
\pagestyle{plain}
\renewcommand{\arraystretch}{1.4}
\hyphenation{stateless}

\begin{abstract}
  Several popular best-practice manifestos for IT design and
  architecture use terms like `stateful', `stateless', `shared
  nothing', etc, and describe `fact based' or `functional'
  descriptions of causal evolution to describe computer processes,
  especially in cloud computing. The concepts are used ambiguously and
  sometimes in contradictory ways, which has led to many imprecise
  beliefs about their implications.  This paper outlines the simple
  view of state and causation in Promise Theory, which accounts for
  the scaling of processes and the relativity of different observers
  in a natural way. It's shown that the concepts of statefulness or
  statelessness are artifacts of observational scale and causal bias
  towards functional evaluation. If we include feedback loops,
  recursion, and process convergence, which appear acausal to
  external observers, the arguments about (im)mutable state need to be
  modified in a scale-dependent way.  In most cases the intended focus
  of such remarks is not terms like `statelessness' but process
  predictability. A simple principle may be substituted in most cases
  as a guide to system design: the principle the separation of dynamic
  scales.

  Understanding data reliance and the ability to keep stable promises
  is of crucial importance to the consistency of data pipelines, and
  distributed client-server interactions, albeit in different ways.
  With increasingly data intensive processes over widely separated
  distributed deployments, e.g. in the Internet of Things and AI
  applications, the effects of instability need a more careful
  treatment. 

  These notes are part of an initiative to engage with thinkers and
  practitioners towards a more rational and disciplined language for
  systems engineering for era of ubiquitous extended-cloud computing.
\end{abstract}

\tableofcontents


\section{Introduction} 

Scaling and reliability of functional systems is a popular topic, not
least for distributed systems.  In Software Engineering, slogans,
manifestos, and best practice frameworks dominate this discussion, and
the academic work on the subject is sparse and has not kept up with
technology. Amongst those slogans, terms like `stateless architecture'
and `immutable infrastructure'
have come to be used to describe a design pattern for software
processes\cite{12factor}, and it is even advocated by some influencers as a principle,
especially in cloud computing\footnote{The terminology
  is part of a pattern of `responsibility free' computing that now
  includes `serverless' as a pattern for consuming a resource without
  responsibility for its underlying dependencies.}.  This is not
without controversy\cite{jonas1,jonas2}, and the lack of agreement
about what it means may well be due to its casual usage: what does
`state' refer to (state of what), and what exactly is the application
versus its platform infrastructure?  A quick search reveals that
there are various definitions of statelessness, all informal, and
usually entangled with specific use-cases. The problems therefore
begin when a usage in one domain spills across into another, bringing
confusion. Counter-proposals involve their own terms and slogans, 
with about the same level of rigour, and eventually these become
quasi-religious convictions rather than rational strategies.

This note is an attempt to disentangle some of these ideas and outline
a reference model that could outlast more than a single generation of
technologies and practices. Some of the issues have been discussed before
in \cite{spacetime2}. The concise summary is that the terms
`stateless', `immutable', and so on, are largely scapegoats for a number of other concepts
that fall under the headings of `dependency', `reliability', and
`fault propagation'. However, I believe that the concepts of locality, state, and
causality are the actual essential ingredients to understand. A few authors
have attempted to explain parts of these issues in the past, but
invariably partially in the course of advocating a particular
recommended practice, so the audience is left with an incomplete
understanding.

The outline of the paper is as follows: in \ref{review} I review a
few rhetorical statements in the popular literature to set up the context for
the discussion.  I explain how no process can be truly stateless, so
we need to understand what authors means when they use `stateless' in
a rhetorical sense. I explain how the observability of state is scale
dependent, and determines boundaries whose partitioning changes the
semantics of promises at different scales.  The important scale is the
one at which an observer assesses the system (this is often a `client'
in a computing setting). In \ref{causality} I discuss the implications
of state in causal determinism, as this is mixed into what authors are
trying to explain.  I extend the usual `past causes future' functional
view of causality to include concurrent, asynchronous, and processes
with feedback, convergent semantics, and desired end states, which are
classically acausal on a the microscopic scale.  Finally, to complete
the popular manifestos, I briefly talk about continuity and
reproducibility (replaying congruent causal sequences) and the
implications of partitioning (modularity) strategy.  Fault domains are a
common idea, but often argued incorrectly.  I try to restate some of
the popular claims to make them more formally correct, and explain why
their original statement is flawed.

Given the scope of the audience, and the importance of reaching as
many readers as possible, my goal is to err on the side of pedagogy
and keep the paper as non-technical as possible---without devolving
into unjustified opinionation. I shall try to provide just
enough justification within the semi-formal language of promises.

\section{Notation and definitions}

In line with previous work\cite{observability,william2,treatise2}, I'll
use the language of Promise Theory\cite{promisebook} to describe
system interactions at a high level.  In a promise theoretic model,
any system is a collection of agents.  Usually, agents will be {\em active
  processes}. Agents represent internalized processes that can make
and keep generalized promises to one another\cite{promisebook}.

The generic label for agents in Promise Theory is $A_i$, where Latin
subscripts $i,j,k,\ldots$ numbers distinguishable agents for
convenience (these effectively become coordinates for the agents).  We shall often
use the symbols $S_i$ and $R_j$, instead, for agents to emphasize
their roles as source (initiator) and receiver (reactive).
So the schematic flow of reasoning is:
\begin{enumerate}
\item $S$ offers (+ promises) data.
\item $R$ accepts (-) promises or rejects the data, either in full or in part.
\item $R$ observes and forms an assessment $\alpha_R(.)$ of what it receives.
\end{enumerate}

\section{Informal ideas about state and causality}\label{review}

The meaning of state can be pursued on many levels.  Suffice it to say
that no decision process or computation can proceed without an
interior dependence on some kind of state\cite{lewis1}. State
basically refers to any information that characterizes a process, over
some interior timescale, and may be stored anywhere within a hierarchy
of agents and subagents that characterize the process. For example, a
clock is a process that maintains an interior state counter. 

The state concept therefore spans the full pantheon of memory storage,
from what is kept in the registers of chips, to configuration files,
source code, or to long term databases---but no two authors will
necessarily agree on which states are the relevant ones to their
arguments, or why they choose to treat one kind of state differently
to another.

\subsection{The role of scale}

Scale plays a role in localizing state. Data may be localized to a
geographical region, a datacentre, a host, a container, a function, or
even a register.  Some authors play the game of offloading state from
one location to another in order to claim statelessness; but that
isn't a scale invariant characteristic.  

I shall argue that genuine characteristics of a system are those that
can be described as invariant properties---i.e. ideas that are not
demolished by a simple change of perspective.  In a virtualized world
of cloud computing, the meaning of being `within' a process, entity,
or agent is ambiguous, because we may draw the
system boundary almost anywhere to focus on specific issues or to
capture extent---and, across the many articles written about
statelessness, there is little agreement about what storage level one
should be talking about. 

\begin{figure}[ht]
\begin{center}
\includegraphics[width=6.5cm]{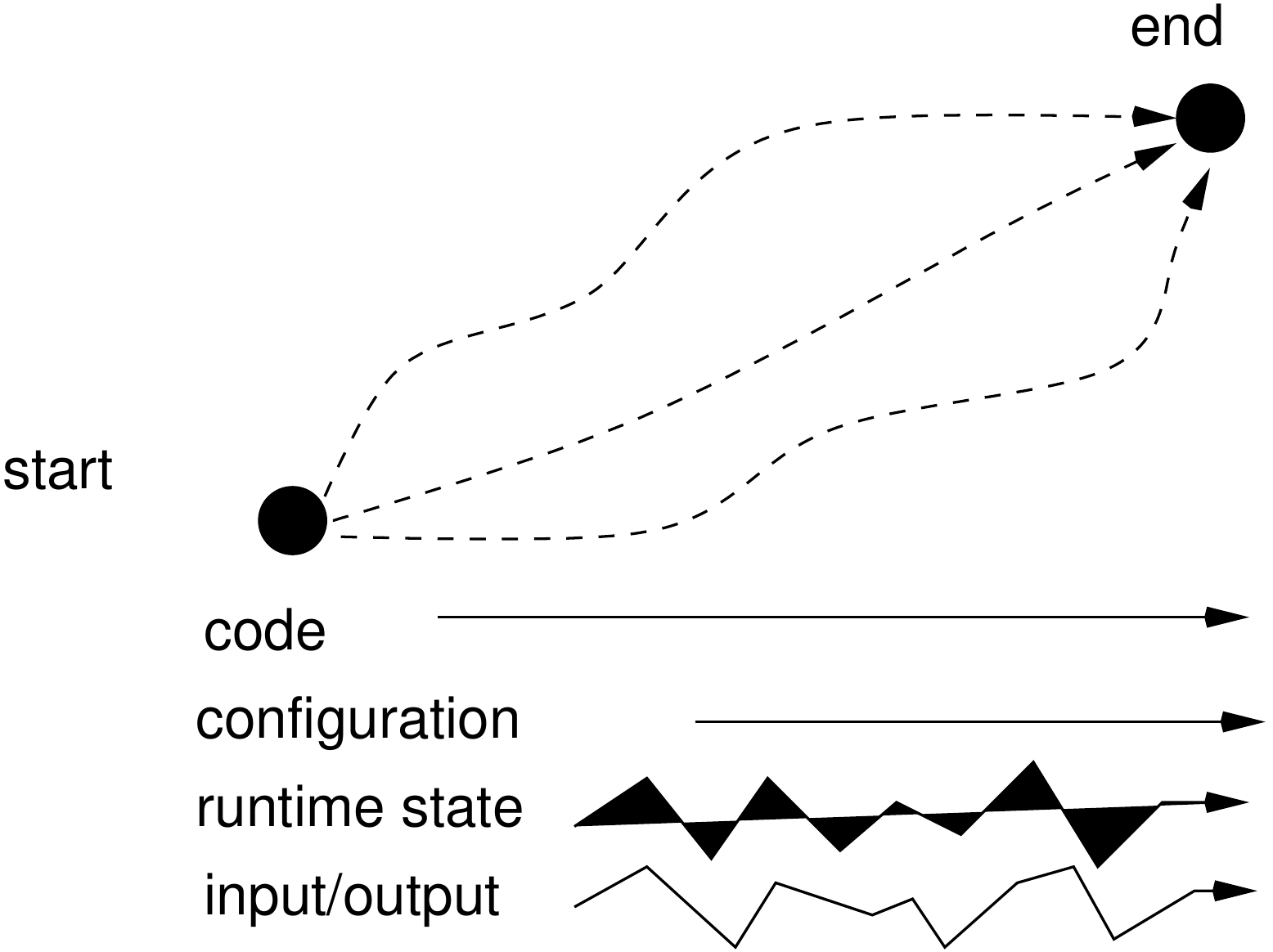}
\caption{\small Processes exhibit state on all manner of timescales.
Some state is frozen into initial state of the packaging, some is allowed to change.
The main question is, over what timescale (or part of the process)
does the state remain invariant?\label{invariance}}
\end{center}
\end{figure}

One author may call a process `stateless' or
`immutable', meaning that all decisions except unavoidable
input-output should be based on state that is frozen and held
invariant before the specific execution of the process (see figure
\ref{invariance}). The scale-dependence of state management, over
space and time, was also the origin of the so-called `configuration
management
wars'\cite{burgessC11,lisa98181,lisa0299,lisa03115,lisa03129}.
Another reader may consider prepackaged frozen configuration choices to belong to a
phase of the processing itself, on a larger timescale, and the lifetime of a single
process is further part of a longer meta-process, involving many
clients, in which continuous delivery of upgrades to changes of
dependencies are interleaved with the keeping of client promises.
Authors thus cherry-pick the meaning of state to suit their arguments.

It's especially important to revisit the topic of state in cloud
computing, where some definitions concerning locality need to be
reconsidered in a scale invariant way; cloud processes often scale
elastically to some extent. Moreover, virtualization adds layers to
its meaning: from source code, configuration, container packaging,
runtime environment, virtual machine, physical host, etc. Developers
constantly jump between the concerns of different levels: from programming,
to continuous delivery, `DevOps', configuration management, serverless, etc.

Various perspectives on these issues have been expressed over the
years\cite{seltzer1,moors1,helland1}.  The Twelve Factor
App\cite{12factor} is a widely referred to best practice manifesto,
which advocates that developers should execute applications as `one or
more stateless processes', and that such apps ought to have a `share
nothing' architecture to avoid contention.  Recommendations then go on
to explain how necessary state can still be kept, after all, by
employing `backing services', and how caching of certain objects is
`allowed'.  This is confusing to say the least. Processes should be
stateless only when we say so? There should be a simple invariant principle.

Nevertheless, there is a hint of a suggestion in the principle of
favouring transactional rather than continuous processing, for a
particular scale and meaning of `transaction'.  So-called `sticky
sessions' that tie multiple web transactions to a particular server
context and location are explicitly rejected in \cite{12factor}.
However, if one takes an extended session to mean a `complete'
dialogue over a business process, including reliable TCP and TLS
negotiations, etc, then it's no longer clear that `stateless', as
implied, has an unambiguous meaning.

Some platforms, like Kubernetes\cite{kubernetes}, have been designed with a notion of
statelessness in mind, but later extended their models to include
state.  Newer additions, such as `service mesh' and sidecars, act as state managers
on behalf of processes and propagate state to reintegrate weakly
coupled systems in a stronger manner.
This suggests that the absence of state itself is not the real problem the
guidelines are clawing at, but that the rejection of what is perceived
as stateful behaviour is really an attempt to address concerns about
localization (scale), speed (timescales), and fault tolerance (spread
prevention). All of this needs to be scaled to cope with the extended
cloud of ubiquitous embedded devices.

\subsection{Popular ideas}

A quick online search and query reveals a number of definitions about
statelessness, which point a finger at state but discuss reliability.
These definitions are generally tied to a single case,
and generalize only by implication\footnote{I choose not to cite the
  ephemeral sources for these `quotes' as they are easily found,
  sometimes paraphrased, and pulled out of longer discussions. I hope
  readers agree that they are representative of the state of
  thinking.}.  For example:
\begin{quote}
\small
  `When an application is stateless, the server does not store any
  state about the client session. Instead, the session data is stored
  on the client and passed to the server as needed.'
\end{quote}
This definition refers to client and server roles in a two-agent
interaction. Similarly:
\begin{quote}
\small
  `A stateful service keeps state `between the connections' or session
  interactions, whereas a stateless service does not'
\end{quote}
In this case `between the connections' is intended to mean that access
to a service is transactional and that some data persist between
independent transactions when a process is stateful.  
The preoccupation with connection indicates the author assumes
a client-server style application.
Wikipedia talks
only about stateless applications, which it defines to mean:
\begin{quote}
\small
  `that no session information is retained by the receiver'
\end{quote}
The substitution of `connections' for `transaction' belies a focus on
client-server computing at a particular scale, and the assumption that
single exchanges are safely invariant while longer exchanges are not.
That would depend on the extent to which the composition of exchanges
were `locked' (e.g. mutex locks) and the data could go missing in case
of interruption.  The excerpt also distinguishes the roles of sender
and receiver as part of the concept, as for a protocol, implying a
directional arrow from client to server.  In general developers tend
to focus their thinking on the preferred scale of the subtask they are
working on, even as ideas about DevOps and Continuous Delivery ask
them to rethink those ideas on a larger scale, for development continuity.

\begin{quote}
\small
`Think of stateless as if a service is a hardware chip. All computation needs
short term storage like registers and stack and maybe heap. What
happens when we lose power? A service that calculates some value and
returns a result can be considered purely stateless.  Purely stateful
would be a service that maintains state like a game server tracking
scores and players in a game world.'
\end{quote}
In this view, scale plays a key role. A short lifetime for data (as
measured in the proper time of the process\footnote{For a definition of
  proper time, see \cite{observability}}, rather than wall-clock
time) means stateless and persistent and reusable means stateful. As
I'll show later, this view of stateless approximates the idea of
memoryless systems (section \ref{memoryless}), and very long term data
that are `invariant' over the effective lifetime of the process can be
separated out and treated differently (see section
\ref{invariantsec}).  A refinement of this:
\begin{quote}
\small
`\item Stateful means written to localhost\\
\item Semi-stateful means 1:1 write over the network\\
\item Semi-stateless means n+1 networks relay eventual write\\
\item Stateless means held in memory until dereferenced.'
\end{quote}
The scale dependence becomes more evident here.  The reference to `in
memory' suggests a short lived once-only usage of state, versus
persistent storage again. The reference to networking is less clear:
which network are we referring to? The interior host bus is a network
alongside the LAN/WAN. In the past, the preference for the processor
bus was about relative speeds: interior communication was much faster
than LAN/WAN communication.  Today, it is impossible to know whether
interior host bus or exterior LAN/WAN connection will be faster.  The
goal of avoiding an architecture that relies on a network connection,
to disk storage, or to a remote service, therfore doesn't stand
scrutiny in the cloud era, as even an in-memory process memory might
be paged out to disk, or retained in a hash table for extended usage.
With this in mind, where exactly is the imagined line between runtime
state and persistent state in the architecture?

\subsection{Process history, entropy, and timescales}

Lamport was the probably the first author to appreciate the relativity
of time in computer science, as a succession of causally ordered
events\cite{vectorclocks}.  The transmission of messages, carrying
causal influence plays a central role in understanding what happens in
computation, both locally and in a distributed system.  Processes that
depend only on a current local register set, i.e. not on the recent
past or the extended history of all such sets are called {\em path
  independent} or {\em memoryless} (see appendix)\footnote{I sometimes
  call them ballistic processes, because we tend to treat computation
  as if it behaves something like a game of billiards: the mere
  sending of some data provokes an immediate involuntary reaction,
  fully deterministic. This is used to argue for `push' over `pull'
  methods, for instance, and is completely wrong.}.  This concept will
be most useful to explaining what authors are trying to express in
`stateless'.

Predictive systems can never be memoryless, for instance, because they
explicitly use past experiences---not only current state---to predict
the near future, involving a computation over multiple samples
collected over multiple proper times. Weather modelling is the
archetypal case in point.  Small differences in the data sets can lead
to large changes in the predictions.  The dependence of data
processing on history is utterly susceptible to scaling arguments.
Similarly, convergent systems that move towards a fixed point or
attractor, at a future time, rely on memory to recognize their final desired end state.

It's hard to generalize about the role of causality, because it is so
dependent on the nature of interactions in a system, but I need to
make a few comments on this because the integrity of data sources has
been challenged in some commentaries, e.g. \cite{helland2,jonas2}.
Some authors have argued that we should never throw away data about past
events because
it might be needed later to `recover' from some fault. Apart from
being unsustainable\footnote{The argument goes: disk storage is
  cheap today, so why wouldn't you store everything forever? There are
  plenty of reasons. For example the escalating power cost of storage
  alone is a reason. Our current experience with the crisis over cheap
  plastics should be a wake up call for anyone advocating an end to
  garbage collection.}, the premise of this argument is wrong from a
causal perspective; indeed, systems must eventually forget their past
over some timescale. The question, again, reduces to understanding the
relevant timescales. Sometimes regulatory bodies insist on data
retention, for legal reasons, up to some statute of limitations. This
can be factored into the policy and separated from the data that
need to accessed dynamically at runtime, becoming effectively a part of
a different application.

Does it matter whether we take data transaction by transaction, or in
bulk as a sum of all transactions (such as an aggregate database or a
file)? The process by which data arrived in the past is only relevant
if it affects the promises it makes at the time of usage by another
agent.  For some authors, a `database' is merely a cache for a long
linear process of accreting the past, i.e. that the current state of a database
is the sum of all past facts transacted at its entry
point\cite{helland2,jonas2}. This is `retarded conditional' view of data: by
integrating differential elements from some beginning to some end one
calculates the answer transaction by transaction.
\begin{quote}
  \small `From this perspective, the contents of the database hold a
  caching of the latest record values in the logs. The truth is the
  log. The database is a cache of a subset of the log. That cached
  subset happens to be the latest value of each record and index value
  from the log.'\cite{helland2}
\end{quote}
This quote summarizes the retarded view of data in which the current
state is merely an arbitrary point is a deterministic trajectory---a
classic Turing machine argument. In this view, all the causal
information lies in the past. 
The argument goes: the present is a function that
does not alter the past (which is true). It's then assumed that the
function is a linear function, formed from a sequence of `deltas' or
state changes that can be stored in a log and added together to yield
the current state. The latter is only true for reversible linearized
(memoryless) processes, localized to a single point of entry. Popular
techniques of this include the use of `actors', `pure functions', and
even mutex locks, with associated costs (i.e.  without interior reads
or writes to exterior data sources)\footnote{See also the approach to
  network data consistency taken in \cite{paulentangle}.}.
Neither of these assumptions generalizes
to distributed cloud computing, as I'll show in section
\ref{causality}. There is also an `advanced conditional view' of
determining outcome, which is also know as `desired end state', in which
future states are the relevant driver of behaviour. 

\begin{example}
  The functional programming manifesto claims that programs will be
  deterministic and reproducible if functions are defined as following
  the following axioms: (i) if they are `total functions', in the
  mathematical sense (i.e. they return an output for every input), ii)
  if they are deterministic, i.e. they return the same output for the
  same input (which assumes they do not implicitly rely on variant
  configuration, database lookups, and are immune to `noise' over all
  timescales involved in the process), and iii) if they alter no
  exterior state other than computing their promised output. The
  composition of such objects would certainly be deterministic, but
  the axioms are often violated in practice, e.g. by system faults and
  by inattention to environmental noise.  The naive view is that
  programs are perfectly isolated and that, if programmers do nothing,
  nothing will happen.  In practice, there is no such isolation
  context for distributed systems, and it's up to programmers to
  explicitly perform noise correction fast enough to maintain these axioms.
\end{example}

It feeds the non-relativistic view that one can absolutely capture
`facts' about the source of information, which can then be preserved
and treated as immutable. The error in this argument is that, as soon
as a sample of data has been transported into storage, it is no longer
the source view: its the observer view of the data store.  One would
have to transport all relevant context into the data snapshot.  This
may be a simple discriminator, but it's still an arbitrary view that
doesn't remove the uncertainties and doesn't warrant the preservation
of an expired context without an understanding of its significance.

In order to recover a snapshot of state, it's argued that one should never
delete any of the contributing facts in the logs. After all, in a
linear system, the current snapshot is merely the balance of all
previous transactions within the system; but this is simplistic. In a
non-linear system, there is no such separation of process timescales,
and we would still need the full past history including all leakages of
noise and interleaved processes to understand the present in general,
because computations are not always linearizable (see section
\ref{linearity}). The final outcome becomes strongly dependent on the
particular moment at which data were collected (a kind of `butterfly
effect')---so both the current snapshot of the database contains
information that is not in the journal\footnote{I suspect that the underlying and unspoken aim of advocating
`stateless' and `throw away nothing' approaches is actually to
linearize systems and make them as deterministic as possible by weak
coupling. Alas, the rising cost of this, in some cases, is prohibitive
and ultimately unsustainable, so alternative strategies should
probably be considered.}.

Even if our system is linear, and we keep all data in an eternal
timeseries, searching backwards takes time, so we index data, but to
do so imposes a rising cost (in energy and labelling), possibly
identifying unique instances, by GUID or quasi-universal timestamps,
and so on.  There is a reason we aggregate data and use caches and
latest summaries: to localize relevant context, and separate it from
other data whose meaning has gone into the mix of entropy.  What every
system designer needs to consider carefully is the extent to which we
flatten a dynamical process into a timeless, static database model.
It's okay to do so as long as you accept the loss of a relative
temporal reference, and the accumulation of entropy.  It does not
necessarily imply that the past is lost.  Process time information is
lost to entropy by design in most systems---and this is not wrong;
relational databases focus entirely on static semantics of data, not
on process histories.  We now have a pantheon of time-series databases
that focus entirely on brute force history, without attention to
`scaled semantics'. By this, I mean that timeseries typically involve
many pattern scales, such as by hour, by week, by month, etc, and that
these are treated as issues for post hoc analysis rather than being
built into the data model in an efficient manner.  There is a policy
choice---a choice of semantics that can't be stipulated in general. If
we want to get systems to behave `properly' as well as efficiently, we
need to select an appropriate causal policy for what is `proper', and
be aware that these choices are inherently scale dependent in both
space and time.

\subsection{The point of usage}

Before summarizing, I want to make one more point about the importance
of the recipient in the determination of so-called facts. Facts are a
kind of promise, but it takes two agents and two promises to pass on
facts: a sender and a receiver.  Past facts are therefore not really
as immutable as is often claimed.  There is the original value offered by
a source $S$ to a receiver $R$:
\beq S \promise{+V} R \eeq and there is the moment at which the value
is accepted and used: \beq R \promise{-V} S.  \eeq It is this latter promise that
actually passes on the information and creates an
event\cite{observability}. We ask: can the promises be kept
invariantly?  The conditions for agent $R$ might have changed, between
the keeping of these two promises, even if $S$ is somehow etched in
stone. Promise Theory predicts that it's not the time of origin of the
data that matters to its causal influence, but rather the moment at
which the data are accepted into the timeline of the next agent---just
as in an electric circuit with feedback, which is the inspiration for
control theory.

If one dabbles in synchronous versus asynchronous processing, this may
matter: if the timescale over which the behaviour of an agent changes
is comparable to the timescale over which you sample data, the data
basically become random variables, by the receiver's hand (not the
source's).  Sufficient immutability can be assured in a few ways: e.g.
by assembling the states one promises to depend on before processing,
to decouple independent processes. That way the processes can continue
at their own rate and still avoid such issues\footnote{This idea was
  built into the design of CFEngine, a realtime maintenance tool
  as a safety measure, and was rarely understood by users.}.  This is
what functions do in programming: automatic variables (by value) copy
the value into private workspace.

There are two approaches: trusting the source and trusting the
receiver.  
\begin{itemize}
\item If one trusts the promise of invariance (immutability) of the
  source $S$ (as one does in timeseries databases, as a trusted second
  hand source), then keeping state there becomes a policy choice and
  one reads information directly from that source.  This second hand
  information replaces the source of `truth', and is not the
  same.  This assumes that there is also an invariant key for looking
  up the data, which is understood by both parties and that
  relationship is also constant.

\item If one trusts the receiver to sample and keep the information
  invariant (as one does in using private local variables in
  programming, and in `immutable images' in cloud computing) then
  policy abhors reading any new information from outside the boundary
  of containment.  Derivative processes, like that of $R$, which
  depend on data from a source $S$, thus capture all their
  dependencies before beginning to keep any subsequent
  promises---freezing them and rendering them immutable as a matter of
  policy.

\end{itemize}
It seems that, in neither case is there any guarantee of the invariance of
the data, or the ability to replay the same interactions multiple times,
as that is entirely a policy decision for $R$. In general data come
from many different sources, with conditions that are quite unequal,
and merely sampling these into a trusted repository does not alter that.
In fact it adds a second layer of trust, by the Intermediate Agent Law (see 7.2.2 of \cite{promisebook}).

What matters is not whether we cache data in a database, or keep each 
update in a journal. What matters is whether the data can be relied upon
not to change over the course of trying to use them.
This often assumes implicitly that there is a single correct
dependency value for each moment in time, with an ability to `roll
back', yet this notion has been debunked\cite{jan60}.

Coarse graining time and separating interior from exterior time: this
is what we do in functional programming. It introduces the full range
of process causal viewpoints.

\begin{quote}
`Mutable state needs to be contained.'
\end{quote}

There is a causal twist here, in the form of Nyquist's sampling law,
and Shannon's error correction law\cite{shannon1,cover1} (for a
review, see \cite{smartspacetime}).  If a system has knowledge of a
correct state (where correct is promised as a matter of policy), then
no unintended deviations from that state will be measured by an
observer if restored quickly enough. We take for granted that such
feedback processes are on-going at a low level of memory in all our
technology at all times.  The same principles may also be applied at a
higher level, as maintenance procedures\footnote{The relationship with
  the strategy used by CFEngine to define `convergent
  operators'\cite{burgessC1,burgessC11}, is interesting.  If you deal
  with pure functions, you cannot have maintenance of persistent
  agents.  You redefine maintenance as the death and rebirth of an
  agent, with associated loss of runtime state. Runtime state is
  contained mutable state. But for a memory process it affects the
  behaviour of the agent in `realtime', i.e. in band of the function's
  I/O channel.}.  If problems are fixed before a fault can be sampled
downstream, there will be no propagation---and the system will be
invariant by virtue of dynamic equilibrium\cite{certainty}.  This is
how data consensus works and memory error correction work, for instance.
The key issue is whether the relevant user can observe any change
in the state or not.

The Twelve Factor App manifesto claims to avoid software erosion, which 
implies that there is should be a maintenance process at work.
In the various cloud manifestos, there has been a focus on
reorganizing the maintenance process so that dependent information is
embedded and assumed invariant (as in a transaction), by freezing
`golden images'. If errors are detected post hoc, due to state drift, one
deletes the process, replacing it with a fresh copy (see the car
example below), which is accepted as a matter of policy. Corrective
actions post hoc (instead of preventative actions) then require some
kind of `rollback' on the scale of the promised transaction.  

Without a preventative error correction underlying the use of runtime
state based on a `fixed image', a post hoc correction is still needed
in that dynamical runtime portion of state, but if the error has been observed by
any process, its influence will already be too late.  In a kernel or
database monitor, keeping validating transactions is relatively easy
(given deep memory error correction), but as the scale of interactions
grows (e.g. in microservices and service meshes), isolation becomes less likely.

\subsection{The importance of forgetting---indistinguishability}\label{memoryless}

Dependency on process history adds baggage (process
mass\cite{smartspacetime}), tangling up changes in dependencies with
data going back in time.  Memoryless processes are `agile' or `cheap'
to run because they have little baggage. This allows changes to be
made to them easily. Of course, that should not be taken to mean that
all changes will be simply localized, with no effect on other
processes.

Keeping processes agile and independent of history seems to be a way
to address quick reproducibility.  Reproducibility has nothing to do with
computation unless it requires replaying an entire journal
of transactions to achieve it. It has more to do with trust in the
stability of promises.  We build businesses and
institutions on reproducibility, because it allows anyone to verify a
result and repair possible errors, when something is judged to go
wrong. At such a time, the idea is that we can forget a state of the
system we consider erroneous, and replace it with a `proper'
policy-acceptable one.

If a process is interrupted, and some of the contributing past
information is lost, we believe that this must compromise the
reproducibility of the outcome. This is not necessarily true, as
explained below, but let's continue.  The result is that we make
transactions that carry all relevant data bundled with them, and keep
a copy until the transaction has successfully been prosecuted.  If the
transaction should fail to be confirmed, we can repeat it.  The
implicit assumption here is that there will be no effect on either
agent (the receiver or the source) unless the transaction completes
successfully. Repetition should also be `safe', i.e. convergent to a
definite outcome, not just a `first come first serve' (FCFS) in a random
walk.

If there are errors on multiple scales, we may have to go back
across cumulative transactions on multiple scales to repeat
the transaction. So, if one builds systems at scale on the basis
of transactional determinism, we are doomed to keeping ever growing amounts of
data, up to the size of the largest transaction. The cost grows in relation
to history (time) not in relation to scale of parallel instances (space).

If any data are left `floating in limbo', in extended `stateful'
sessions, it is argued that those states could be lost and data may go
missing.  This is not about statefulness at the receiver, but about when the data are
discarded from the source so that the receiver state can no longer
be reproduced. This naively seems to return us to a justification
for the idea of never throwing away any data, discussed above, but
this is not so. We simply need to preserve data until confirmation of
receipt---as in reliable transfer protocols. In other words: we
must assess when data have already played their role in the next
stage of the computational pipeline.  Next we need to define
the scale of that remark: on what process scale do we need
confirmation of `ok to delete'? If we treat transactions as packet by
packet over a session, then a process crash could lose data.  But if
we treat completion as the confirmation of a promise kept that depends
on the data, then scaled transactions can be constructed using locks.

Safety under repetition is the much neglected method of assuring
certainty in systems. Idempotence is sometimes mentioned, but most
authors think this means remembering which transactions are completed
on a FCFS basis without checking for contradictions.  

\begin{example}
  Numbering of transactions, like in TCP, is one way to maintain
  coherence of order, but this is not always meaningful without ad hoc
  assumptions.  In a data pipeline, for example, you can number items,
  but the numbers assume that both ends have a clear sense of how the
  arrival of data will take place in order to combine multiple sources
  meaningfully.  So, while a 1:N transport can be regularized by
  partial ordering, N:1 aggregation cannot.  Numbering promises
  process as `intentional' events, but random arrivals have no such
  coherence, making the processes non-reproducible unless the entire
  history is captured and used as the future source of truth. That
  assumed truth may not be a faithful representation of the original
  source processes. As always, the receiver determines the semantics
of data.
\end{example}
Fixed point
convergence is the more economical key to reproducibility, because a fixed point is the
only certain way of guiding a system with random behaviour to a known
state. It requires knowledge of a future state to which the system is
headed.  Absolute invariant future state is simple and cheap to
manage.  Relative future state is fragile and susceptible to faults
and cumulative errors of execution.  In data pipelines, for instance,
this needs to be treated very carefully (see our Koalja work, for
instance\cite{koalja}).

\begin{example}[State of a car]
  If you crash your car, the car will be lost if it is a unique,
  one-of-a-kind design. But if the memory of its design (its ideal
  state, minus age and mileage decrepitation) is kept elsewhere, as a
  separate manufacturing process, then the car can be replaced---but
  not its runtime state, i.e. the precise details of what it was doing
  at the time of the crash (including its passenger inventory).
If the car client is not fussy, it may overlook a few details and be
satisfied with an equivalent.
\end{example}
Some of the state you are happy to forget, some you are attached to.
There is no fact present in the car that can tell you how to
discriminate this line.  We don't always need to remember the past,
sometimes only a single future `desired' state.  Indeed, it's
desirable to forget the past as it's just in the way---and sends you a
regular bill.

\begin{example}
A statute of limitations, or causality horizon. If you build on advanced boundary
conditions, you need no memory. Memoryless processes have a very short horizon.
A function is basically a mutex lock around a private cache of function argument
values. The completion of the function is a tick of its clock at the scale of functions,
and therefore a pure function is memoryless at that scale.
\end{example}

We might conclude that favouring
`statelessness' is to push responsibility for preserving
state backwards along a causal chain (into the past): onto the sources
rather than the receivers. But this is not necessarily true. Trajectories
also depend on policy of future states and the rules of transaction, and there is
significant freedom to redefine causal constraints by shifting responsibility
between these, over different timescales.
We have to ask: which of the agents in this chain is the most
fragile? Why should events from the past be more important or `correct' than
what happens in the future? Many will answer that `the past determines
the future', but this naive determinism. Promises about future states
also contain causal information, and it is not correct once one accounts properly for
scaling. So we need to return to look at the ways in which causal
propagation takes place and is scale dependent.

\subsection{Summary: proper invariants}

To summarize, we appear to have succumbed to the trap of obsessing over illusory
detail rather than focusing on the key question: how can a stable promise be
kept?  So what we seem to be struggling to express is a design
decision (a promise) about which processes will be considered atomic
at each scale, which is equivalent to expressing which data we are
potentially willing to lose.

\begin{assumption}[Promise manifesto]
  \small The central question about systems is: will the outcome of
  promises be invariant to the conditions under which the promise is
  kept or not---and does that matter to the promisee?\footnote{The recipient
may have arranged for contingencies in advance.}
\end{assumption}

From this perspective, statelessness actually seems to imply a preference
to use `current state' in short-lived, ephemeral interactions, over
which dependencies can be treated as approximate invariants.  If all agent
interactions are kept short, as measured by their own proper time, and
relative to the scale of their larger process's exterior time, 
\beq 
\frac{\Delta t_\text{interior}}{\Delta t_\text{exterior}} &\ll& 1  \label{ts}\\
~\nonumber\\
~\nonumber\\
\text{i.e.  } ~~~~~ \frac{\Delta t_\text{transaction}}{\Delta t_\text{process}} &\ll& 1. 
\eeq 
then a
process will tend to a state of statistical invariance.  This relative
timescale argument could perhaps be used as a definition of `micro' in
microservice.  It makes dynamical sense: it's a linearization of a
potentially non-linear process. We should understand that as a
design constraint. The choice enables eventually consistent outcomes over 
a sample set, but there may be other sample sets that
have not reached the equilibrium. The best promise (no guarantee)
of stability is to ensure that updates
have plenty of time to reach equilibrium, by separating timescales. 

\begin{example}
  If a dependency changes every second, and a process promises output
  every few seconds, there is insufficient time for the process to
  promise invariance. However, if changes to dependencies occur only
  once per year, then processes lasting a few seconds can be
  considered invariant in practice, by (\ref{ts}).
\end{example}

There is an implicit separation of concerns in talking about state:
the part of state that we care about, in the current context, and the
part we don't. This suggests a natural partitioning {\em by policy} of
scales for each relative process, rather than a universal best
practice guideline.  The final point about `good enough replacement'
leads us to consider the role of observability and distinguishability
in deciding outcomes\cite{observability}.

The issue in question seems to be: over what timescale can some form
of state be considered dependable (invariant relative to the
receiver), from the perspective of all stakeholders in the system?
This includes at least the role of the client (when data are uploaded)
and the server (receiver of uploaded data).  For the remainder of the
paper, I'll therefore focus on the dynamical principles of keeping
promises across a multitude of scales.

\section{Locality and distinguishability}\label{locality}

The focus of the paper, from here on, will be to illustrate how a few key
concepts behind the promises authors try to capture in
rhetorical usage may be formalized. The central concepts are mainly spacetime concepts, about
order, scale, and observation\cite{observability}.

\subsection{Localization or spatial partitioning}

The virtue of source code modularity, for the separation of {\em
  semantic} concerns, is doctrine in computer science.  Localization
of process execution in space is a form of modularity too, that we
call scaling of execution context---`containment' for short. Today,
virtual machines and container technologies are the tools for
achieving such spatial localization, erecting barriers that are
supposed to limit the exchange of influence between interior and
exterior.  Isolation from an influence $X$ implies causal independence
of $X$ (see section \ref{cause}). In Promise Theory, bare agents that
make no promises are assumed independent of all causal influence a
priori.

\begin{figure}[ht]
\begin{center}
\includegraphics[width=6.5cm]{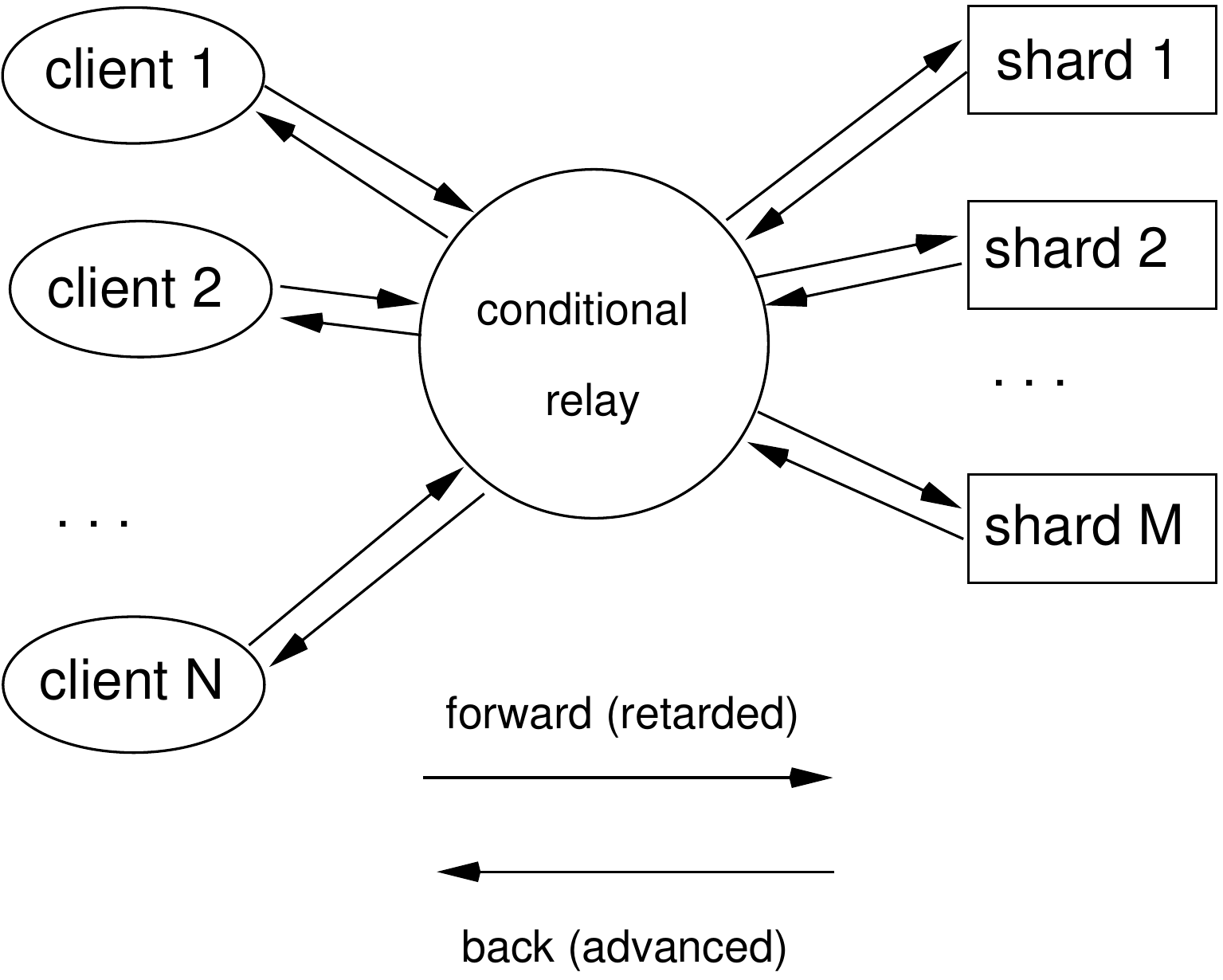}
\caption{\small Causal influence can flow forwards or backwards along
  the direction of an interaction (defined arbitrarily). At the ends
  of an $N:M$ interaction, say from clients to servers, data may be
  shared or kept in entirely separate scopes, in either direction.
Where nothing is shared, we often say that the data are `sharded' or have 
private scope. Integrating shard implies reintroducing a shared resource in
the agent that aggregates them however, so we don't escape sharing; we
only delay its onset in order to acquire partial invariance of
data for other processes.
\label{advret}}
\end{center}
\end{figure}
In Promise Theory, every active or passive part of a system is an agent.
The definition of an agent also defines a scale, and an isolation
boundary for that scale.  Elementary agents are the smallest
observable scale of a system, and superagent clusters of them form
larger scales, where agents work together to keep collaborative promises.

Modularity is achieved by partitioning the process into separate
agents whose interactions are defined by promises.
\begin{definition}[Partitioning of a process]
  \small A subdivision of a process agent into a number of
  non-overlapping subagents, in such a way that the mutual promises
  between the subagents are exposed to exterior observers.
\end{definition}
Agents may be decomposed by space or by time if they are distinguishable
by some label. In other words, if agents
are numbered in order, or labelled with names or types, they can be separated
into subdivisions using the labels they promise.  Examples include the
division of a larger process into microservices, or the partitioning
of a database into shards, perhaps curated by intermediaries with a
APIs between them, but the principle doesn't refer to any particular
technology or set of assumptions.

Now, let's formalize the hierarchy of agents involved in representing data processing,
starting with the easy parts. This helps to establish the language of promises
and use of terminology.

\subsection{Scaling of state}

Every memory location in a system that can record state is an agent
that can promise to hold a value\footnote{This applies to
the nodes in any state machine too.}.  All states are memory agents, and
partitionings lead to separation of states that keep different
promises---sometimes called `sharding' (see figures \ref{advret} and \ref{hierarchy2}):
\begin{figure}[ht]
\begin{center}
\includegraphics[width=5.5cm]{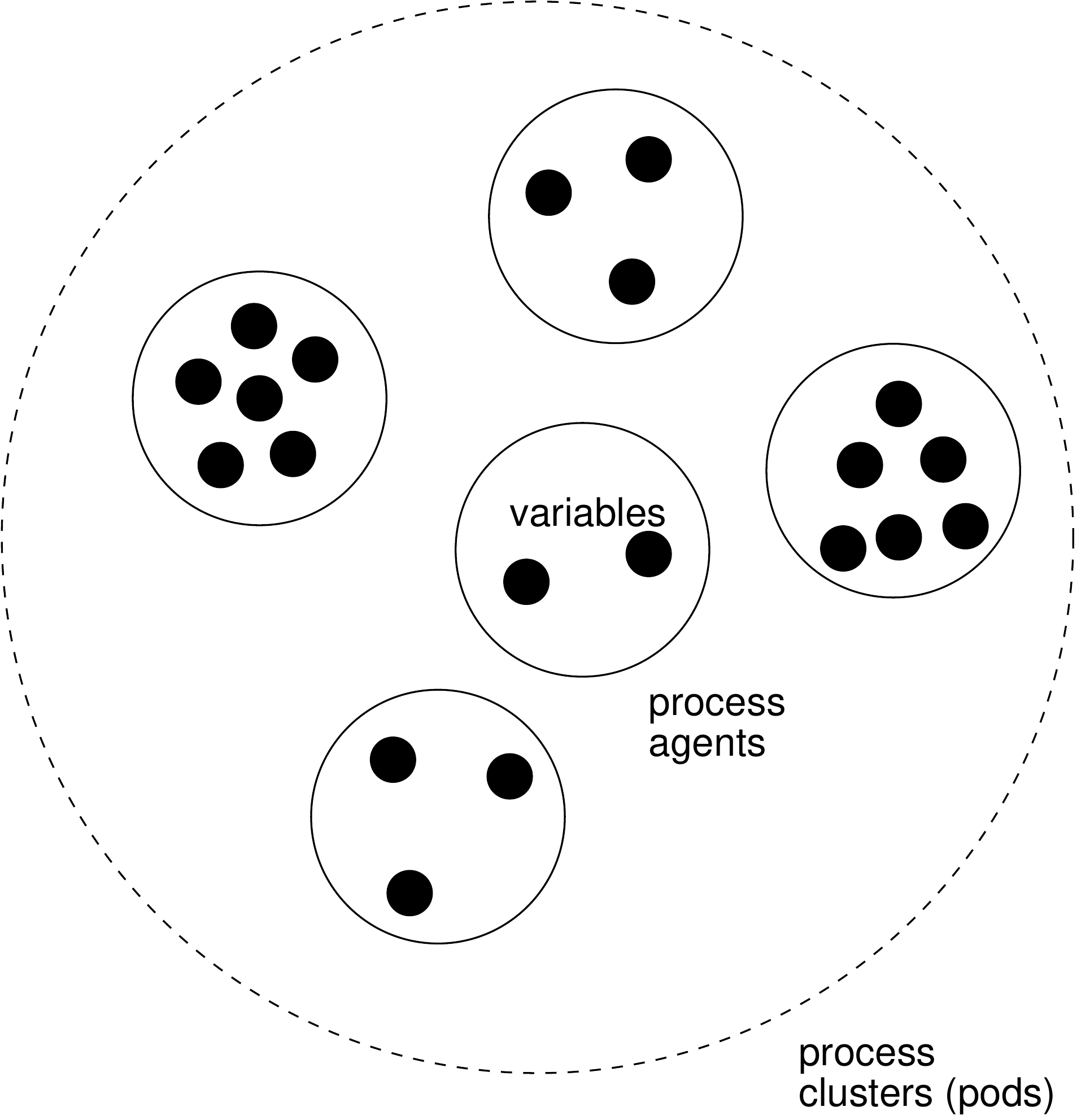}
\caption{\small Distinguishability of agents to clients determines whether they can
be partitioned or whether they form a redundant set. Agents are distinguished
by the promises they make, which in turn are states of the agent.
Together the states of a system form a configuration.
Some states are promised to exterior agents and
  some have private scope. In general the scope of a promised
  value is contained by a certain scale, which we call a semantic
  boundary. Redundant agents are indistinguishable. Non-redundant
shards make different promises.\label{hierarchy2}}
\end{center}
\end{figure}

\begin{definition}[Variable]
\small
A agent or subagent $V$ promising a key-value pair, that promises a name and a value, representable as
a simple agent:
promise.
\beq
V \promise{+\text{(name,value)}} S,
\eeq
within a scope $S$.
\end{definition}
The internal variables of a process agent are what one normally thinks of
as the state of the agent.
\begin{definition}[State of a variable]\label{statedef}
\small
  Let $V$ be any variable (or set of variables), on the interior of a
  process agent $S$, which takes values from any set of
  distinguishable elements $X$. The state of $V$ is the value of $V
  \in X$, promised by the source $S$ to any outside agent $\Unspec$: 
\beq 
S \promise{+V} \Unspec.  
\eeq
\end{definition}
The problem with this definition is that the state is only observable to the
process agent $O$ outside of $A$, if both the source $S$ promises the information
as an exterior promise, and the
observer also promises to accept and use the promised information:
\begin{definition}[Observable state of a variable]\label{statedef2}
\small
Let the state of a variable $V_S \in X$, promised by $S$,
be sampled on any occasion by an observer $O$:
\beq
S \promise{+V_S} O\\
O \promise{-V_O} S,
\eeq
where $V_O \in X$. Then the observable state of the variable is $V_S \;\intersection\; V_O$.
\end{definition}
The role played by the observer in this definition is crucial. It
underlines how relativity will play a role in all stateful
phenomena\footnote{The example of observing the inconsistent state of
  a clock was discussed in reference \cite{observability}.}.  If a
state is persistent or invariant to order $N$ samples, then multiple samples, by an
observer $O$, lead to the same value for some number of samples $N$.
It is clear that a variable, even of order 1 implies the existence of
memory on that can be sampled by some process that carries information
to an observer. State is an observer issue rather than a provider issue.

\begin{definition}[State of an agent]
\small The total state of an agent consists of the states of its interior variables.
\end{definition}
In effect, each variable is a subagent member of a larger process
(superagent).  This tells us that the boundary where we choose to
define the edge of a process plays an important role in the way we
describe its behaviours (process, container, group, pod, host, etc).
Moreover, since it relates to promises, whatever else it may be, the
statefulness of an agent $S$ is an {\em assessment} made by each
recipient observer involved in promised interactions.

\subsection{Sequences or temporal partitioning}

There is also localization in time: when a process's trajectory starts
and ends (see figure \ref{invariance}), and whether information is fed
into it only at those endpoints as immutable constants or `invariants'
of the process, or whether information is accepted into it and
modifies the process as it evolves.  The purpose of `functions' in
programming is to promise that only the I/O channels belonging to the
function (the arguments and the return value) lead to change. This is
hard to assure on a larger scale, however, as computer code is only
one in a mixture of overlapping processes whose dependencies lead to
mixing.  Seeking invariance of promises is the key to process
stability. A process may be called {\em adiabatic}\cite{burgesstheory}
if exterior information does not alter promise definitions over the
timescale of interactions that rely on it---meaning that a process's
promises are invariant over the interval during which they are being
kept, with no configuration changes\footnote{Confusion ensues for many
  when considering the origins for such change.  There may be
  intentional change, such as a code change or a manual input of data,
  and there may be unintentional (hidden) change to a dependency
  presumed invariant. A lot of rhetoric has been exchanged around
  `never touch the system and it will never go wrong', but `if it
  fails, don't fix it---replace it'. These are policy decisions, not
  unique recipes for handling change, but they are rooted in the idea
  that invariance is a solid foundation for process continuity.  They
  may have different causal outcomes.}.  If process fragments are
partitioned end to end, they form a sequence.  If they coexist, either
starting or ending at a common agent at a common time, then they may
be called concurrent.

Localization allows us to partition processes in space and time, 
holding certain aspects constant over the duration of a sub-process.
\begin{itemize}
\item Time localization leads to promise invariance for agents.

\item Space locality leads to privacy of scope and non-interference.
\end{itemize}
This is a form of lock-free synchronization. Proponents of the Actor Model
will find these principles familiar\cite{actormodel}.

\subsection{Distinguishability, partitions, and redundancy}

In order to distinguish 
partitioned agents from redundant agents, partitions must be distinguishable by the agents
that interact with them.

\begin{definition}[Redundant agents]
\small
Two agents $A_1$ and $A_2$ are observationally redundant if they make the same
promises to an observer $A_3$, and the $A_3$ accepts the promises equally, i.e.
\beq
A_1 \promise{+X} A_3\\
A_2 \promise{+X} A_3\\
A_3 \promise{-Y} A_1\\
A_3 \promise{-Y} A_2
\eeq
where $X \intersection Y\not=\emptyset$.
\end{definition}
An observer that discriminates between two agents making identical
promises may be called a discriminator. Such agents are the basis of
all decision making based on data. In principle, discrimination at a
single agent location can be made in on the basis of space (source
agent) or arrival time, but each simultaneous time step sampled by the
discriminator represents a new causal decision, so that must depend on how
we define the scale timesteps and the discriminator itself---a distributed
superagent discriminator has to be able to promise interior time coherence.
As always, the scale of encapsulation over which we can assume
invariance (`coherence') plays the main complicating role.

\begin{definition}[Partitioned agents]
\small
Let two collections of agents $P_1=\{A_1, \ldots\}$ and $P_2=\{A_2, \ldots\}$ 
be partitions, as discriminated by an observer $A_3$, then:
\beq
P_1 \promise{+X_1} A_3\\
P_2 \promise{+X_2} A_3\\
A_3 \promise{-Y} A_1\\
A_3 \promise{-Y} A_2
\eeq
where $X_1 \intersection Y\not=\emptyset$, $X_2 \intersection Y\not=\emptyset$, and $X_1 \intersection X_2 = \emptyset$.
\end{definition}
Note that a partitioning is a superagent, i.e. 
\begin{lemma}[Partitions are agents]
If agents $A_i$ are scale $n$, then
a partition is a superagent at scale $n+1$.
\end{lemma}
This follows trivially from the definitions, without any restrictions on what other promises
the agents may make.

\begin{example}
  `Shared nothing' agents cannot be completely redundant. They make
  {\em a priori} uncorrelated and uncalibrated promises, so they
  cannot promise determined redundancy; they are merely random and
  possibly similar. As soon as they accept a common source of
  calibration (by cooperative dependency on a single source), or
  achieve dynamical equilibration (as in data consistency protocols), they
  share something from $O(1)$ to $O(N^2)$. In practice, the claim of
  `shared nothing' is overtstated as agents share determining configuration
  that only changes on a long timescale, making it effectively
  invariant, according to (\ref{ts}).
\end{example}

\subsection{Invariance}\label{invariantsec}

We can now define the meaning of {\em invariance} of an interaction, as a process involving
pairs of agents (on any scale): a source and an observer:

\begin{definition}[Invariance of a promise $\pi$]
\small
An agent's (exterior) promise may be called invariant
if the body of the promise is constant for the lifetime of the
promise.
\end{definition}

\begin{theorem}[Invariant promises]
\small
  An agent may be invariant with respect to exterior change if and
  only if it contains all its dependencies and promises them constant.
\end{theorem}
To prove this, suppose an agent $A$ makes a promise of $X$ to an observer $O$, conditionally on the
promise of another agent $A_D$ being kept, which promises a dependency $D$, then:
\beq
\pi_A: A &\promise{X | D}& O\\
\pi_{A^T}: A &\promise{-D}& A_D\\
\pi_D: D &\promise{+D}& A.
\eeq
In addition, $A$ promises its value of $D$ to be constant:
\beq
\pi_{C}: A &\promise{D=\text{const}}& O
\eeq
If we form the superagent from $\{ A, A_D\}$, from the two collaborating agents, then
\beq
\{ A, A_D\} &\promise{+X | D}& O,\\
\{ A, A_D\} &\promise{-D=\text{const}}& O,
\eeq
which is unconditional after $A$ assesses the promise to provide $D$ has been kept, i.e. $\alpha_A(\pi_D)\not=0$.
Now, since $D$ is promised constant, $X|D \rightarrow X|\text{const}$, and this is invariant under $D$.
If the promise $\pi_A$ is made unconditionally, then $D=\emptyset$, and the result is trivially true.

\begin{example}
  \small This theorem may be considered the basis for freezing all
  dependencies and configurations internally in containers before
  execution, e.g. in Docker or using fixed images in the cloud.  But
  it also applies to dynamical configuration engines, like CFEngine
  etc, where the policy for $D$ is fixed and the promise keeping is
  maintained dynamically by `self-healing' based on fixed policy. In
  either case, the promise may be broken because the result cannot be
  guaranteed.  In the case of containment, one is trusting the
  integrity of the containment, which can only be assured for
  non-runtime state by making it read-only. In the case of dynamical
  configuration, runtime state can also be repaired by dynamical
  equilibrium in the presence of noise.  So the required invariance is
  not dependent on a particular strategy.  A dynamical configuration
  is more expensive in processing, but may prevent errors before they
  occur. Static containment may appear cheap, in terms of runtime
  process resources, but is more likely to result in exterior
  consequences that breach containment, because the timescale of
  exposure to non-corrective actions is maximal (the lifetime of the
  process) rather than a regular shorter maintenance interval.
\end{example}
We can reduce this to a very simple expression of invariance for agents, as a whole:
\begin{definition}[Invariance of an agent $A$]
\small
An agent promises to accept nothing from any agent.
\end{definition}
From the proof, we see that this is a scale dependent assertion, since
we may always partition the agent internally such that one interior
partition makes a promise on which the other interior partition
depends, entirely within the boundary of the agent, leaving its
exterior promise unconditional.

The key assumption in this argument is the absence of unintended change, by impositions,
such as noise, that systems are fragile to. Many developers believe
that there is no noise in systems, only the programmed change, because
a lot of it has been eliminated by low level error correction\footnote{In band
`self healing' configuration engines are essentially noise error correction processes, on
a fairly long timescale of minutes to hours, which may be too slow to maintain
invariance for busy processes.}.

As long as `a system' of choice interacts with some other agency, it
is not the total system, merely an arbitrary partitioning of it. If an
promises to accept nothing, i.e. make no (-) promises, then its
interior state will be invariant for as long as that promise can be
kept. We may assume that this is the actual goal of systems that serve users.

\subsection{Sharing versus partitioning}

Partitioning is naturally the opposite of sharing.  The original definition
defined `shared nothing' for databases was `neither memory nor
peripheral storage is shared among processors'\cite{sharednothing}. In
the cloud era, we need a more generalized abstraction to cope with the
branching technologies.
\begin{definition}[`Shared nothing' agent]
\small
An agent is keeps all of its promises unconditionally (makes no assisted promises),
from its own intrinsic capabilities, i.e. it makes no promises that require
the assistance of another agent.
\end{definition}
An example would be a unikernel architecture without network service
interactions and private disk.  Since `shared nothing' the default
assumption of `autonomous' behaviour for agents in Promise Theory, we
see the utility of promises to describing these issues: every
dependency has to be revealed as a promise to see the channels that
constrain process operation. A simple consequence of defining this is
that agents that play the mediating roles of hubs, switches, or
routers (as in figure \ref{advret}) for promises of any kind violate the condition above. 

\begin{lemma}[Hubs violate `shared nothing']
\small
Any nodes that operate as a point of confluence, or a divergence like a switch,
and connect a sharding of process messages, promises
to partake in sharing and violates the assumptions of `shared nothing' in the broader sense.
Shared nothing involves promises that do not depend on one another for any of
their resources. This even includes power supply at the deepest level.
\end{lemma}
The degree of sensitivity to sharing depends on the possible variance of the 
dependency. If we seek to depend only on invariants, then there is only weak
coupling. The shorter the timescale for variation (the more active a dependency
is), the greater its effect on the system promises.

The connection between state and partitioning lies in what information
is used to distinguish process agents. Process trajectories trace the
evolution of causal relationships from agent to agent, at whatever
scale an observer can witness.  Some agents may make
indistinguishable promises leading to {\em redundant parallelism}, or
they can promise full distinguishability leading to {\em branching}
and {\em switching} (decision making).  

Distinguishability of promises is what enables non-shared futures,
i.e. sharding and switching (see figure \ref{advret}).  In switching, a process selects from a
set of possible futures based on the state of variable data.  Each
decision partitions possible outcomes into branches, or `many worlds'
futures.  If branches are indistinguishable (contain only the same
redundant information, both in initial conditions and runtime state)
then the branching process is memoryless, and the superposition of
agents acts as a single superagent on a larger scale. If they are
distinguishable, the program takes on a new course.

Occasionally, different process flows merge into a single one. This
happens with pull requests in software development, for example.  It
also happens in data pipelines where source information gets
aggregated into batches.  Branching (+ promises) costs nothing, but
merging timelines (- promises) requires causal intervention, and the
input of new information in the form of state-dependent selection
criteria.  Thus we do not escape the cost of a memory process by
branching as long as there is a need for the branches to be
merged\footnote{This is basically the reason why Continuous Delivery
  advocates recommend developing software in a single branch.
  Contention can then be resolved in band, since software development
  is a largely stateful process, in spite of modularity.}.

\begin{example}
\small
  In network package delivery, i.e. `routing', for instance, the
  decision about which route to take is variable according to a
  separate parallel process, but that might be made on the same
  timescale as the running process from which the data arises. This is
  then non-linear (see section \ref{linearity}).  It is always downstream
  (receiver) promises that carry the greatest responsibility and the
  greatest potential cost---hence the downstream principle (see
  section \ref{downstreamsec}).
\end{example}

\section{States, events, and messages}

State refers to information, which may be relied upon by some process
to determine its next steps. This includes monitoring systems that may
rely on the state to trigger exterior processes.  The total state of
the world is all the {\em observable} information in the universe
(including chains of indirect dependency).  That which cannot be
observed by an agent cannot influence its future.  Information may be
hosted by agents in the form of any promisable properties, and overlap
between human and computer subsystems, for instance as part of a
larger business process.  Partial state may localized to a particular
container, or be aggregated over those locations (multiple agents), as
well as over time (multiple episodes)\footnote{These two methods
  correspond to frequency (space) and Bayesian (time) interpretations
  of statistical state.}.

States get passed between agents and their processes, coupling them
together.  When a sample of data arrives, it normally leads to a
transition from one state to another\cite{lewis1}. This is the meaning of a message.

\begin{definition}[Message]
\small
A discrete unit of data, i.e. a state transition carried between agents.
\end{definition}
A message channel is a pair of promises:
\beq
S \promise{+M_S} R\\
R \promise{-M_R} S,
\eeq
where $M_S \intersection M_R \not= \emptyset$, that forms a non-empty promise binding to share messages unidirectionally from
a sender to a receiver.

We understand events as `happenings'. In physics, we attribute
coordinates to events, in space and time; this is the origin of many
confusions. In computer science, coordinates refer to process
signposts and program counters, which are not generally helpful to
know externally (their scope is local)\cite{observability}. The key
difference between a message an an event is that events are
observational in character.
\begin{definition}[Event]
\small
A discrete unit of process in which an atomic state change is observed or sampled.
\end{definition}
We further imagine processes being driven by a flow of events, like a
stream; that's because observers serialize them as a matter of policy,
based on the limitations of their cognitive processes---and situate themselves
downstream of the outcomes they are interested in.
A message only becomes an event
when it is sampled (i.e. accepted by a receiver) and generates a change
of state, becoming a tick in the clock of that process.
\begin{definition}[Event or Message Driven Agent]
\small
Any agent $R$ that can promise the occurrence of an event $E$, conditionally on its sampling
of message $M$ from a source $S$, with an average rate $\lambda$:
\beq
R \imposition{+E | M} O\label{emp}
\eeq
i.e. $R$ can promise an observer $O$ that it acknowledges an event $E$ on receipt
of a message $M$. By Promise Theory axioms, this assumes the prior promises (or impositions):
\beq
S \imposition{+M | \lambda} R &\text{~~\rm or~~}& S \promise{+M | \lambda} R\\
R &\promise{-M | \mu}& S
\eeq
where $\mu$ is the queue service rate.
\end{definition}
A message is the transmission of a proposed state change. If accepted
there is a response on the interior of the agent, and there may or may not be
a response on the exterior. This includes heartbeats and repetitions of all kinds;
no prejudice should be inferred what is important and non-important state.

\subsection{Propagation of influence by state messages}\label{propag}

We see that state does not lead to influence unless it is observed,
and there are conditional promises that use it to promise conditional
behaviour. This is the behaviour of a `switch'. 

We can now think about this in terms of dependencies. In order for remote state to
affect a local process, a source agent has to share it, then the receiver
has to observe, accept it, and subsequently alter its behaviour
according to it (see section \ref{linearity}). 

From section \ref{locality}, what we call the beginning and end of a
process is a scale-dependent characterization. We make a choice about
which agents we want to include at a given scale. The common understanding
of processes has some basic elements however. Each process has a lifecycle
of major states that characterize it, which we can call epochs in the lifecycle
of the process.
\begin{itemize}
\item Definition (of promises).
\item Initialization of resources (Initial state).
\item Execution (keeping promises).
\item Termination (Final State).
\end{itemize}
This is basically the same model one has of any dynamical system in
mathematics or physics. It maps on to the equivalent, e.g. think
of solving differential equations:
\begin{itemize}
\item Definition the equation.
\item Initial boundary condition.
\item Find the propagator that computes the derivative states.
\item Final boundary condition.
\end{itemize}
These are the major elements we use to describe the causality
of a system, and fix its trajectory.

In a cloud setting, these correspond to
\begin{itemize}
\item Building software
\item Configuring software settings
\item Executing the software (runtime)
\item The desired end state (outcome)
\end{itemize}
There is state in each of these epoch timescales. 
We are free to redefine the placement of changes. e.g. keeping code
or configuration invariant during execution, or to write code or configuration
that rewrites itself (as in learning systems).

\subsection{Scaling of local state}

A proper discussion about the localization of state can only made with
reference to a theory of scaling. What is local at one scale is
composed of many locations on a smaller scale\footnote{The description of a virtual hierarchy of perimeter boundaries around
resources leads to a kind of `Gauss law' for promises made by process
agents.  Any promise of state expressible externally must come from
interior process memory.}?  We therefore need to
decide on the agents, or units of localization: what do we mean by
entity, agent, location in a given context?  As mentioned above, a certain
locale could refer to anything from single chip register or a
distributed database, depending on the author's state of mind!

Computer processes are made up agents, which are discrete processing units.
They sometimes work together in clusters---represented here as superagents.
By making promises, they form many patterns such as client-server interactions, data
pipelines, object models, microservices, container pods, backup
servers, redundant failover, etc.  Promise Theory provides a simple
view of scaling, based on boundary semantics, that easily accounts for
the cases found in IT\cite{spacetime2}.  We can thus ask, to what
extent are promises (e.g. about state) within or without of a
boundary?  Is state implicated in decision-making at the level of a
conditional promise on the interior or exterior of an agent boundary?
How is state implicated in propagation of assisted promise-keeping?

Locality refers then to the ability to draw a semantically defined
boundary around an agent (i.e. one based on what it promises rather
than based on where it happens to reside) and decide what is on its
interior (local) and what is exterior to it (non-local).  Every system
of agents that interacts with other agents breaches its boundary or
grows it to accommodate new members, so the definition of a system
`module' is always an ad hoc matter.  Modules are often chosen based
on functional separation in IT\footnote{I've argued that one should
  instead be guided by {\em The Principle Of Separation Of Timescales}
  if predictability and stability are the primary
  goal\cite{observability,treatise1,treatise2}.}.

Part of the confusion in the colloquial use of `stateless' is that `state' itself
refers an implicit and specific scale for many authors, namely
whatever favoured object they happen to be working on, such as a programming
class, a process container, a cluster, or a host computer, etc. Software engineering
does not teach practitioners to think across multiple scales.
State may therefore refer to all scales, from interior microstates to
aggregate macrostates, and refer to real or virtual space. In order to
observe and measure state, it needs to persist relative to the process
that samples it (i.e. for some finite number of samples or duration of
proper process time). Different processes tick at different rates, and
interactions often lead to waiting. The issues of observation were
discussed in \cite{observability}.

A notion of `total state' may be accumulated over many interactions,
either laterally across many redundant concurrent processes, or
longitudinally over multiple similar interactions, such as in data
collection and machine learning applications.  Already it seems clear
that we need to distinguish different kinds of state and that the
intended use of the data play a part in what is objectionable about
statefulness to some authors. If we think of sampling as an
information channel, in the Shannon sense, then the separation of
timescales amounts to partitioning process samples into different
channels according to the timescales over which we {\em assume} that
certain state we rely on will be invariant, i.e. constant with
respect to multiple samples.

\subsection{State localization at different scales}

If we redefine the agent boundaries or partitions of a system, we can
shift state formally from one location to another, but we can't do so
without altering the promises kept by the outer boundary. This
assumes, naturally, that state is depended on for a purpose. Free state is
irrelevant baggage.

We can try to summarize statelessness without referring to a
particular case, like client-server or data pipeline, or even to a
particular scale, while---at the same time---unifying semantics and dynamics
for the process:
\begin{definition}[Locally stateful]
\small
  A locally stateful process is one in which memory is kept on the
  interior of a process agent or superagent cluster. This memory is
  promised for as long as the process agent's dependent exterior
  promises persist, and access to process memory occurs over interior
  channels.
\end{definition}

\begin{figure}[ht]
\begin{center}
\includegraphics[width=6.5cm]{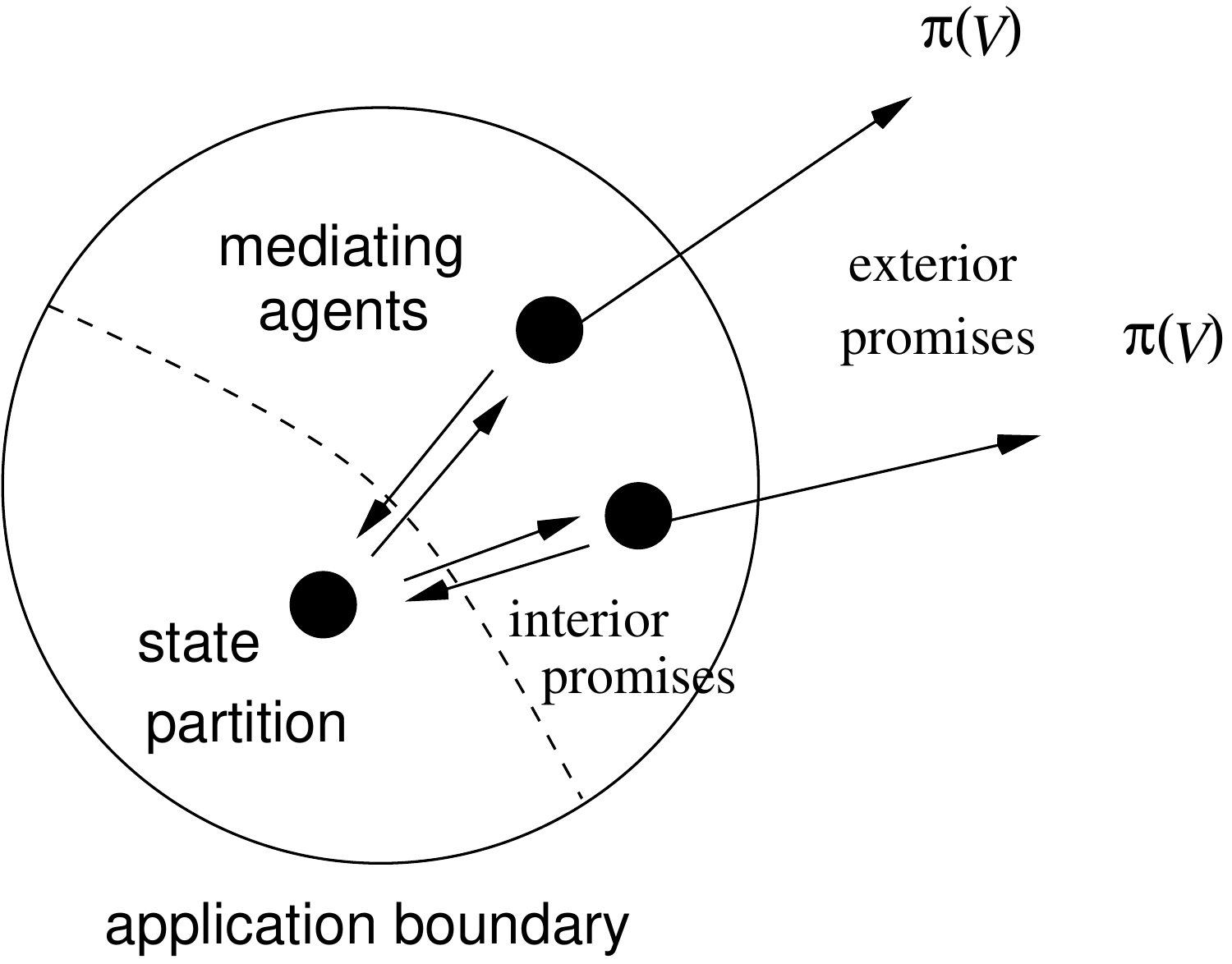}
\caption{\small A partitioning of a process `entity' or superagent.
  Within its semantic boundary there are stateful parts and stateless
  parts. Should we argue these as separate or integrated? The exterior
  promises are conditional on the interior state dependency, but the
  mediating agents are stateless (memoryless).  This approach was used
  in Kubernetes, for example, where container services were initially
  assumed weakly stateless, with possible database services
  partitioned into separate containers and storage services. Later,
  this was rationalized to weaken the claim of
  statelessness.\label{partition}}
\end{center}
\end{figure}

\begin{definition}[Non-locally stateful]
\small
  A non-locally stateful agent is a composite agent, in which any
  persistent process memory accumulated over the history of
  interactions is partitioned and kept independently of the agent
  mediating an exterior conditional promise (see figure \ref{partition}).  The mediating agent is
  then merely a conduit for state that persists in an agent
  belonging to a `backing' process partition.  The loss of the
  mediating agent does not incur a loss of partitioned process memory
  for the collaboration.
\end{definition}
This deliberate indirection---pushing state out of one agent and into
a dependency---seems to implicitly reference shared resources and risk
mitigation, not whether state is promised or used\footnote{The intent
  of the Twelve-Factor App manifesto seems to principally address risk
  and local contention.}.
So reference to state is a red herring for intended purpose, and it
conceals assumptions about the timescales and number of times over
which the state will be used before it changes. Such matters are
critical and therefore the assumptions are unacceptable.
\begin{figure}[ht]
\begin{center}
\includegraphics[width=7.5cm]{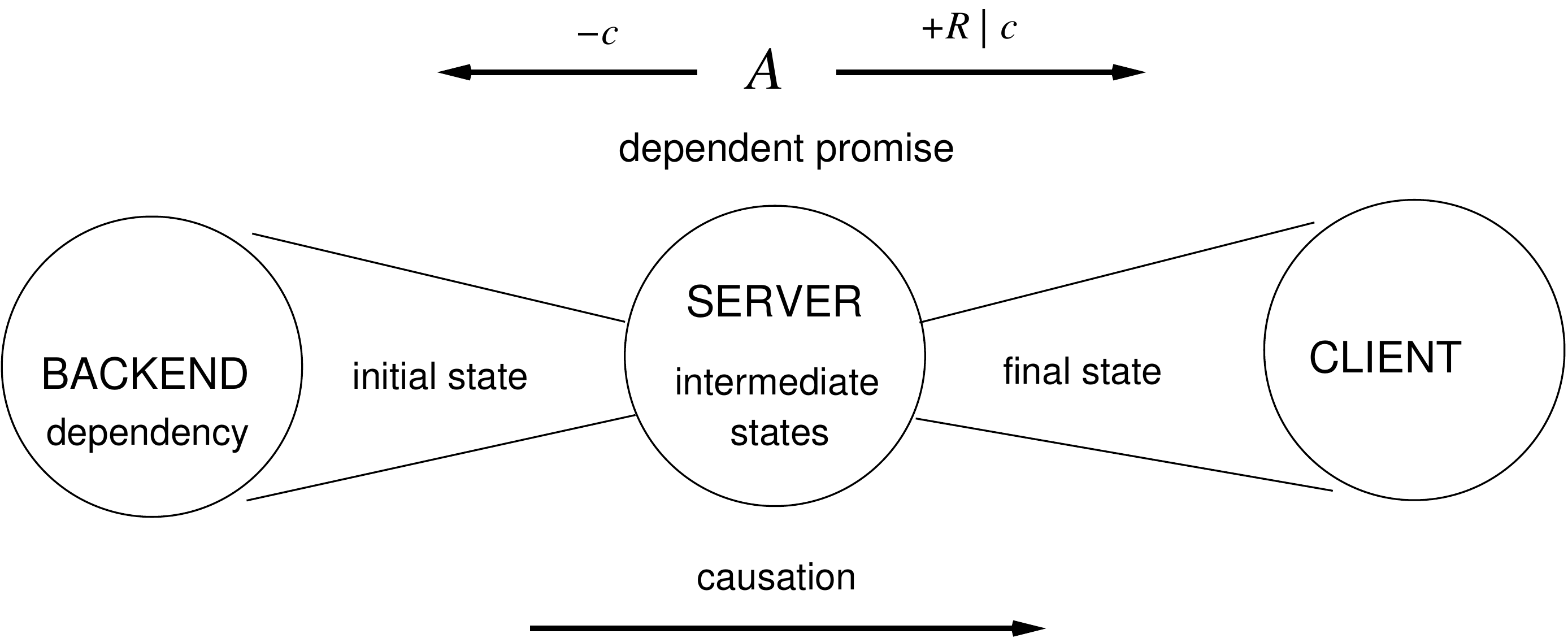}
\caption{\small Conditionality is causality. The states that are
  causally implicated in a conditional promise include initial
  state, including code and prior configuration; then there is runtime
  state and shortlived intermediate computational state, which may be
  reconstructible from code and inputs. All these combine to an output
  that may be strongly or weakly dependent on the full set: how
  many intermediate states are involved in computing a virtual
  transaction? Is the scope of that state private or shared between
  distinct exchanges? These are all questions of process
  scale.\label{scattering}}
\end{center}
\end{figure}

The key question about state, then, is not whether it is retained, but
rather whether or not it is used as a dependency in the keeping of a
larger promise. If the loss or latency of such state gets in the way
of a larger dependent promise being kept, then one would be better
served by a collaborative architecture in which that risk may be
mitigated.  The term `shared nothing architecture'\cite{sharednothing}
is more accurate than `stateless' to address this.  It implies a form
of sharding or partitioning of agency in a system: possibly at either
the client side or the server side.  Both ends can end up having to
deal with inconsistent promises\footnote{A huge amount of discussion
  centres on data consensus sharing protocols for server (+) promises,
  but almost nothing is written about the responsibility of the
  receivers (-) who ultimately shoulder the burden of dealing with
  inconsistency. For an industrial example, see \cite{breck1}.}.

Whether we keep state in primary RAM memory rather than in secondary
disk storage or even tertiary services like databases is not the
issue. The issue is how do we depend on it, i.e. what happens if it's
lost. What sources do we {\em trust} to keep stable promises?

Unfortunately, a `shared nothing' partitioning of dependencies to
localize causal interference has its own problems.  A set of services
(e.g. for web, database, and storage) is already made up of separate
processes, even if they run on the same host, or with the same common
storage.  Just how much separation is `shared nothing'?  If they all
share a common purpose, then they must be connected by something.
Should we wrap them in layers of virtualization (containers, virtual
machines, etc), or run them on different hosts, in different racks, in
different datacentres?  By handing off state to another agent that
serves it up as a backing service, we only introduce a new shared
dependency.

\section{Stateful (memory) processes}

We can now put these key elements together to understand causal
dependence, or chains and transition matrices across networks of
agents. What's key about a process is what invariant information we
have to constrain its trajectory. Initial and final conditions are the
available external fixed points. The process rules (promises) may also
implicitly contain fixed point behaviour such that the process
converges to a `desired state'.

When processes depend on one another, they observe one another's
states. The amount of memory they have, internally, defines the extent
of their dependence on their own causal past or future, both as memory
for storing their program and for representing decisions as a `log' or `journal'
of prior states. 

\subsection{Short memory processes: linearity}\label{linearity}

A perspective, which addresses increasingly popular ideas
about complexity and chaotic behaviours, is process
{\em linearity}.  Linearity is related to weak coupling,
and addresses the relative scales of process interactions (see the
earlier paper on observability \cite{observability}). Non-linearity is
associated with memory behaviour---behaviours in which past
interactions change the system so that new interactions experience a
modified system. This is {\em learning} behaviour. A non-linear process
cannot simply be replaced or restarted without access to a complete
history of interactions, synchronized for all times, because it's
outcomes depend on that unique memory of interactions. Non-linear
agents cannot be redundant, as their unique histories distinguish
them.

A system may be called linear if it comprises conditional promises
that are Markov processes of order no greater than 1 (Markov processes
are described in the appendix). In other words, if the process is
independent of past inputs to a scale that goes back $n>1$ samples
into the past (where the `past' is defined to mean a chain of prior
samples). This matters when the delivery of data could be carried out
in a transactional way, but the promised methods that receive and
process the data are changing concurrently as part of an independent
process.  Dependency graphs may span multiple processes implicitly.
They might be quite invisible in program code.

Consider a simple interaction, of the `client-server' variety, in which
an agent $C$ (in the role of client) promises or imposes a request $c$ onto
an agent $S$ (in the role of server), which is accepted by $S$ i.e. 
\beq
C &\imposition{+c}& S\\
S &\promise{-c}& C. 
\eeq
The absorption of $c$ by $S$ implies that a state has changed in $S$, for some
timescale that persists for a sufficiently long time to enable a response $r$ to be
returned. Let's say that the response is a simple storage lookup, like a database record
or a web page. This acts as a key-value pair, where the key is $c$ and the value is $r(c)$,
which depends on $c$
\beq
S \promise{+r(c)\;|\;c} C.
\eeq
In order for $S$ to make this conditional promise, it has to contain
the state variable $V=r(c)$ on its interior. The state variable is
persistent, so $S$ is clearly part of a system that promises state.
Now, it might `outsource' this capability to another agent (a backing
service, in the vocabulary of \cite{12factor}). Then, we have an
assisted promise\cite{promisebook}. Suppose the assisting or backing
agent is $D$, then $S$ hands off responsibility for state to a
subordinate agent, and must therefore make an assisted promise that
depends both on the client request $c$ and the promise of state
storage $d$:
\beq
S &\promise{+r(c)\;|\;c,d}& C,\\
S &\promise{-d}& D,\\
S &\promise{-c}& C,\\
\eeq
where $S$ hands off the request to its subordinate:
\beq
S &\promise{+c'(c) \;|\; c}& D\\
D &\promise{-c'}&  S\\
D &\promise{+d(c')\; |\; c'}& S\\
S &\promise{-d(c')}& D
\eeq
As long as each dependence is a Markov process, forming a Markov chain, 
the dependency on $c$ is linear.
\begin{definition}[Linear conditional promise]
\small
A conditional promise $\pi$ is linear with respect to a dependency $d$ iff,
\beq
\pi: ~ S \promise{+V(d) \; | \; d} R
\eeq
implies that $\partial V/\partial d = \text{const}$ over the life of $\pi$ (see appendix).
\end{definition}
Linearity literally implies that a functional dependence on $d$
is linear (of polynomial order 1), and does not alter the
functional form of the promise $V(d)$. The dependency $d$ does not
alter the promise, except to act as a lookup key. If we were to repeat
the keeping of the promise over some timescale, i.e. over some chain
of promise keeping assessments, an observer would not assess there to
be any difference in the result of $V(d)$, over a number of samples
$T$. The promise is therefore invariant over a timescale $T$.  

The qualification of a bounded interval $T$ is important, because no
system is truly invariant for all future history (see figure
\ref{invariance}).  Changes do occur to systems and their promises:
new versions of software promises are made, for example. The real
issue is whether one can redefine a process to ensure that invariants
are fixed somehow before runtime execution starts and all the way up
to when it ends\footnote{This is a more precise expression of what
  people mean by `immutable containers'.}.

\begin{lemma}[Linear promises and weak coupling]
\small
The need to wait for state history increases the service time
for a queue, increasing the ratio of $\lambda_R/\lambda_S$.
\end{lemma}
We need to define clear {\em
  timescales} for the assertions (promises) we make.  Slowly varying changes
decouple from changes that occur on the timescale of the promise
because each sampling of a linear system is an independent variable,
and a sequence that depends on multiple samples is independent of the
sample if the sample has already been integrated (e.g. refactored)
into the definition of $\pi$\footnote{This is what happens, for
  example, when runtime incidents lead to iterated bug fixes in
  software and new promises are made that incorporate past states into
  the initial conditions of current promises (feedback loops). The
  process of Continuous Delivery renders such changes on the same
  timescale as runtime transactions, which makes sensitivity to change
  higher.}.

\subsection{Long memory processes}

Long memory processes depend on the sequences of states that led to their
current state: e.g. does it matter which route you used to enter the city?
This is the typical domain of machine learning.

The memory required to keep this promise determines a minimum scale
for the process. Long memory processes cannot be stateless, in any
definition, but it may be possible to separate part of a long memory
process and isolate certain subagents whose behaviour is memoryless.

\subsection{Invariant definitions of stateless}

Given the popular usage of the term `stateless', it seems appropriate
to accommodate the commonplace ideas with a clearer definition, so that we do
least violence to present day intuitions. This leads to what I'll call
weak statelessness:

\begin{definition}[Weakly stateless process]
\small
  A memoryless process (Markov process of order 1) promises that its
  interior memory of past interactions is the empty set: 
\beq A
  \promise{+V(t)=\emptyset} *.
\eeq 
\end{definition}
The definition is only weak, because it doesn't say much
about what other behaviours the process may have.
Implicitly, it suggests that that the next outcome of the process
can only depend on the inputs at each step. Inputs could easily include
data from long term exterior memory. The key point is that the promises
that are purely local to the weakly stateless process are decoupled from, i.e.
invariant, for all possible input-output transitions, as in (\ref{markov}).

Memory processes, or stateful processes, are those
that are not weakly stateless.
\begin{definition}[Stateful (memory) process]
\small
A process that promises:
\beq
A \promise{+V(t)\not=\emptyset} *.
\eeq
\end{definition}
When these two kinds of process are composed, to form a superagent on a larger scale,
the result is naturally stateful.
\begin{lemma}[Stateful + stateless = stateful]
An agent that promises to be both stateful and weakly stateless is stateful
by composition.
\end{lemma}
The proof is trivial:
\beq
\left. 
\begin{array}{c}
A \promise{+\emptyset} *\\
A \promise{+V(t)} *
\end{array}
\right\}~~~ \equiv ~~~A \promise{+V(t)} *
\eeq
If we want to be strict in the definition of statelessness (what we might
call a purely ballistic process) then the agent responsible has to refuse
all input.
\begin{definition}[Strongly stateless process]
\small
  A process that has no exterior (-) promises to accept input from any
  source during its lifetime.  The agent's promise is thus completely
  constant: it does not rely on the order or substance of any other
  information.
\end{definition}

\begin{figure}[ht]
\begin{center}
\includegraphics[width=7.5cm]{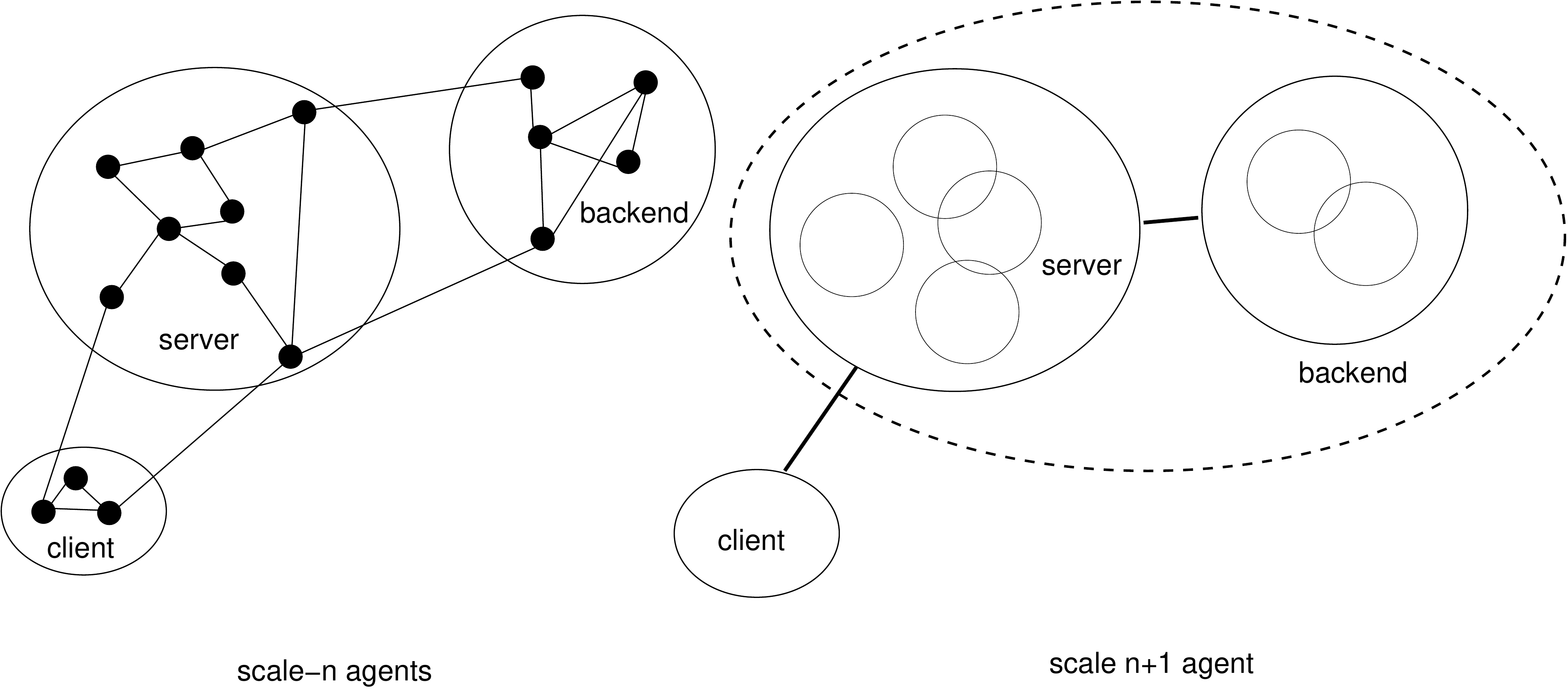}
\caption{\small A client-server system with a backend can treat the backend as part of
a service, or as a separate service. If the exterior promises remain the same, then
these configurations are indistinguishable.
We are always free to compose or decompose agents at scale $n$ into
agents at scale $n-1$ or $n+1$ by redrawing the boundaries around modules.
This shifts a discussion about interior to exterior or vice versa, but cannot
affect the outcome observed by an agent on a scale greater than the total system.\label{scales3}}
\end{center}
\end{figure}
What we surmise is that basically all non-trivial processes must be
stateful on some scale, because a promise of stateful behaviour
overrides a promise of stateless behaviour on any scale.

\subsection{Transactions on scale $T$}

It's usual to define transactions in terms of atomicity and consistency. 
Here we can define the concepts more simply using invariance of promises:
\begin{definition}[Transaction at scale $T$]
\small  A transaction is the promise of an invariant sequence of messages
  $M_1,M_2,\ldots,M_T$, of length/number $T$, accepted by a process
  agent $A$, whose memory of the messages is also invariant over the
  sequence, and contains all the data needed to keep the conditional
  promise \beq A \promise{+X | M_1,M_2,\ldots,M_T} \eeq
\end{definition}
In other words, the agent $A$ doesn't let go of the information 
from its cache until it
is acknowledged by the receiver. Failures on a large enough scale 
can still wipe out all
the information of the transaction, but this adds some assurance of
invariance if the data survive the transaction.

With this definition, we do not presuppose any model or scale for the
meaning of a transaction. As long as the transacting agent is
invariant over the completion of its promised task, and the data
require no dependencies. The virtue of this definition is to make such
transactions repeatable, as all the conditions of the transaction are
self-contained, and thus invariant. Put another way, transactions turn
messages into scalable autonomous (super)agents, without exterior
dependencies beyond their promised scale $T$.

\begin{lemma}[Transactions are repeatable]
\small
Any valid transaction at scale $T$ leads to a repeatable process,
given the same message and conditional promise.
\end{lemma}
Notice that the process is only memoryless if $T=1$, i.e. we choose
a particular scale, but the all important invariance
is scale independent. Also note that it's important to distinguish between
events and transactions, which many authors fail to do. The invariant
properties of transactions are not shared by arbitrary messages, so 
favouring a transactional system is not the same as favouring a message or event
driven system.

\subsection{Scale dependence of state and causality}

Under scaling transformations that aggregate processes by causal dependence,
the foregoing discussion should make it clear that we can state quite
strongly:
\begin{theorem}[Statelessness is scale dependent]
\small
  A process that is weakly stateless at scale $n$ may be stateful when
  causal promises are composed or decomposed at scales $n-1$ and $n+1$.
\end{theorem}
The proof of this is elementary. Consider agents $A_1$ and $A_2$ that
make promises that are stateless and stateful respectively, such that
$A_1$ depends on the promise of $A_2$
\beq
A_1 &\promise{\text{stateless}}& *\\
A_2 &\promise{\text{stateful}}& *\\
\{ A_1, A_2 \}  &\promise{\text{stateful}}& *
\eeq
This theorem renders statements like `transactions cannot span
entities'\cite{helland1} meaningless, as there is no plausible
definition of an entity without a clear specification of scale.

Causality itself is about the transmission of prior state, along the
trajectory of each autonomous process, causality must itself be scale
dependent. Indeed, as we'll see, influences may appear to be
determined by states that are only reached in a process's future. Time
does not follow a simple imperative ballistic view of prior state. In
the frame of the process itself (the proper time) acausal changes
frequently take place, by advanced boundary information. 

\section{Causality and event driven propagation}\label{causality}

Several authors have commented on the importance to time relativity
for understanding process
execution\cite{arewethereyet,jonas3,kiki}---already bringing insights
from spacetime relativity, and `many worlds' interpretations of Kripke
and Everett\cite{kripke1,everett1}.  Time has been the domain of
physics for centuries, and it would be a mistake to not pay attention
to the full range of patterns developed there.  To fully understand
causality in distributed systems we need to expand the simplistic
understanding of universal past, now, future into a local view in
which causal behaviour depends on {\em all three} in a {\em scale
  dependent} way.

\subsection{Past, present, and now}

Past, now, and future are concepts about the order of events relative
to a process of observation.  What an observer calls `now' is the
state expressed by its clock, i.e. a snapshot of its complete interior
state (see interior time\cite{observability}).  Obviously, this is not
a scale invariant assertion---if we step back, or zoom in, the
boundary between interior and exterior is altered\footnote{This is why
  the concept of a microservice architecture versus monolith has no invariant
definition either.}.
Agents may be aggregated into superagents, which are the smallest
grains on a larger scale.

The common view of causation is the retarded view:
\begin{definition}[Retarded process]
\small In a chain of dependent promises,
  a process depends on an invariant initial state or boundary
  condition. The final state of the agent does not play a role in
  determining the outcome of the process.
\end{definition}
\begin{example}
  In a process to build a tower, the balance of the project bank
  account starts with the invariant boundary condition of zero money.
  Its final state is a sum of transactions related to that initial
  state. The final outcome of the tower plays no role in determining
  the final amount in the bank account.
\end{example}
The contrary view, often used in radio engineering is:
\begin{definition}[Advanced process]\small (includes desired end-state,
  recursion, etc).  In a chain of dependent promises, a process
  depends on an invariant final state or boundary condition. The
  initial state of the agent does not play a role in determining the
  outcome of the process.
\end{definition}

\begin{example}
In the space race to the moon, the final invariant outcome of the process
was to land a person on the moon. The chain of transactions leading
to that point was not dependent on the initial conditions of the
project.
\end{example}
In the latter case, the final state of an agent is implicitly or
self-determined, and the promises work backwards to search for a path
to reach it from the undefined initial state. This approach is used in
transactional `rollback', for instance. It's also how a GPS navigation
system works, for example. A process to solve a Rubik's cube is also
anchored in the invariant future state (the desired end state of
ordered colour\cite{kiki}).

\begin{example}[Proper time clocks]
\small
For example, the increments of time for a cloud process could be
measured by ticks that represent the starting and stopping of
container processes.  Or we go count each function call as a tick of a
clock, or each statement.  This is not nit-picking: it matters to the
issue of causality how we define the evolution of progress.
\end{example}
In a flowchart view of programming, which represents the most common 
imperative view of time, the future is thought of as a function of the
past. 
\beq
T_\text{next} = f(T_\text{this}).
\eeq
Each prior statement leads inevitably to the next by an implicit
jump instruction in the process counter.  No statement changes the
past, because everything advances at the same rate: the result of each
statement is the essentially deterministic keeping of an exterior
promise of its agent. 

\begin{figure}[ht]
\begin{center}
\includegraphics[width=7cm]{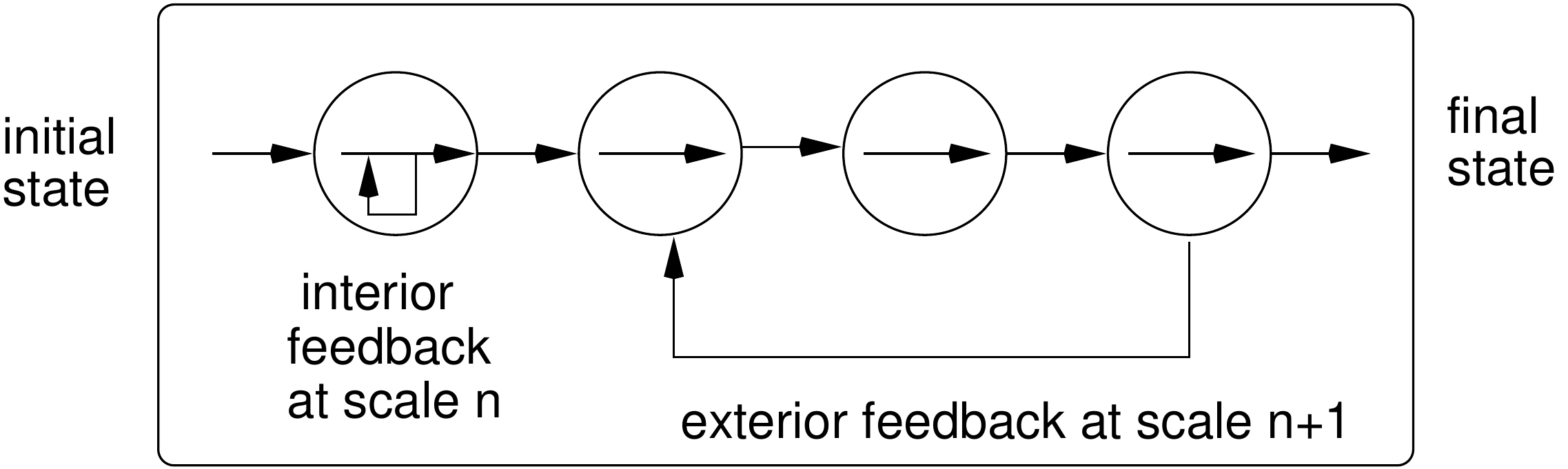}
\caption{\small Feedback on the functional scale, including iteration
  and recursion, leads to apparent causation in the reverse direction
  relative to large scale exterior time (against the observed flow as
  seen by an exterior observer), but because this is unobservable, the
  promised outcome, measured by exterior clock time always appears to
  flow in a constant direction from start to finish. This is what we
  interpret as from past to future. The direction of time on a large
  scale is from left to right, but inside the subroutines it may be
  counter to this monotonic progress.\label{acausal}}
\end{center}
\end{figure}

At a function call level, this is somewhat ambiguous, because a
function call involves recursion, which poses the promise of
the outcome before the execution the keeps the promise, i.e. in the assignment $x := f(y)$,
the right hand side is assumes that $f(y)$ exists, which involves
stack frames to create a sideways dimension of `subtime'\footnote{I
  borrow this phrase from Paul Borrill, and elsewhere use the term
  `interior time'.}, whose incorporation into the process is acausal
from the perspective of a programmer.  Past and future get muddled by
an assignment that behaves like an advanced boundary condition, while
a subroutine advances with a locally retarded boundary condition of
the function argument.  The discrete scaling of a process into lumps,
or subroutines, implies that time does not run in a simple fashion for
any observer outside the system (see figure \ref{acausal}).

Functions that are non-deterministic may also employ data
that are not accounted for by the promises of the agent\cite{jan60}.
Such systems are known to be irreversible, but can be made to behave
consistently by using advanced (exterior future) boundary values.

\begin{definition}[Interior feedback]
\small
Interior feedback at scale $n$ is a causal sequence of messages whose
channel runs counter to the direction of the system's proper time at
scale $n+1$. It is unobservable from the exterior of the agent
containing it. In other words, the process clock only ticks
after iterations and interior subtime machinations have reached
an outcome that can keep the agent's exterior promise.
\end{definition}
Feedback may appear causal or acausal depending on the scale of the
agent making that assessment. This only illustrates how the meaning
of time is naturally complicated in distributed multiscale processes.
It's neither deep nor trivial.
\begin{definition}[Exterior feedback]
\small
Exterior feedback is the same from the perspective on the inside.
A dependency from downstream of the process (the causal future) which
is merged with a dependency from upstream (the causal past).
\end{definition}

\subsection{Facts, messages, and event horizons}

Messages are the transport mechanism for program transitions between
agents. Events are the observation of a state transition by any agent
$O$.  The preservation of an event, as an immutable fact, is not {\em
  a priori} guaranteed by any agent. It is design choice (a promise)
made by the receiver, whose default is non-immutability.  An invariant
promise by an agent needs interior memory to remember it, and---since all resources are
finite, including memory---there must also be a cut-off lifetime for such facts to be
remembered (an event horizon).  As the scale of a dependent process,
it encompasses an increasing amount of memory, which implies a growing
power cost and increased interior time latency for data retrieval.
Eventually, the ability to recall prior facts must become much greater
than the lifetime of the agent's promise lifetime.

\begin{example}
\small
  This cost has been made clear in the early blockchains, where
  coherence or consistency of the chain (the transaction journal) is
  the causal promise.
\end{example}

The idea that the past informs the future is too simplistic for
distributed processes. A model of computation is a model of causally
ordered events, but the order of causality is actually undefined
because we are free to place certain information in the rules of propagation and
other information in the boundary conditions, instead of all in one
place.

Einstein taught us that causality is what an observer sees.  The
arrival of messages, leading to events, defines a perceived direction
for time for each observer independently.  It is always measured at
the scale of whatever observer assesses it.  What happens on the
interior, including acausal feedback loops, is usually discounted (see
figure \ref{acausal}). Interior process sequence numbers may be used
to pay for determinism of causal ordering by coarse graining time,
i.e. by paying for order preservation with a delay in {\em interior
  time}; this may not appear to delay the process on the exterior, but
adds a cost in terms of unique distinguishability of messages to
sender and receiver.  Other information, like desired end states,
fixed points and other `attractors' may bring about convergence around
states that only exist in the relative future, and a process only ends
(in interior time or subtime) when that future state has been
reached\footnote{This is Turing's halting problem.}.

\subsection{Repeatability and fixed points}\label{idemp}

The true goal of information systems as tools is to strive for
repeatability or predictability. It should now be clear that this is
about the larger goal of arranging invariance over process conditions,
i.e. dependencies.  The surrogates that often stand in place of this,
such a statelessness, and causal ordering, are themselves
non-invariant characteristics and should therefore be avoided.

A common mistake is to try to assure invariance by acting `only once' (the FCFS
random walk approach to state, rather than the determined fixed point).
For example, in the delivery of a transaction. We might number transactions,
like TCP sequence numbers, and tick them off a checklist as they are completed.
This leads to a growing process memory (a stateful process). It can be
replaced by a memoryless local process using advanced causation.

Advanced causation (treating the end state as a fixed point) has many
uses, e.g. for desired state policy enforcement.  Systems whose
interior states are changing may not have homogeneous transitions
across different replays, but a choice of a fixed attractor is
equivalent to inline error correction.

Relying on thing that happen only once is a non-invariant procedure
(changing the sampling timeout can change yes into no).
Messages may be repeated or lost, and isolation from interference is not a promise that
can be kept easily (process isolation is often the first thing
violated by intrusions and security exploits). If we seek a deeper
level of safety, it makes sense to rely not on the keeping of promises
that are fed as data, but on the characteristics that are more likely
to be preserved, such as convergence to fixed points\footnote{This
  explains the value of in band configuration maintenance for security
  and safety in a live setting. Instead of relying on isolation, one
  hopes for isolation but validates it with a competing immunity
  process (`trust but verify').}.  The surest means to achieve
repeatability is the maintain the promises on a timescale shorter than
that at which they are sampled. This is the Nyquist sampling theorem
in action.

Advanced propagation determines
based on a desired state $x_D$
\beq
x_\text{end} = f(x_\text{any})\\
x_\text{end} = f(x_\text{end})
\eeq 
We see that the final value is insensitive to the initial value, which
is in strict opposition to the functional idea of past forming immutable facts.
The immutable fact lies in the definition of the function itself, which
refers to an `inevitable' future state.

The outcome is idempotent when it reaches its final state, not after a
certain number of transactions `once only' has been
reached\cite{treatise1,burgessC11}. The approach is what the immune
system does, and was used famously in
CFEngine\cite{burgessC1,burgessC11} and later configuration
tools\footnote{This distinction and its scale dependence was the basis
  for the configuration management wars of the 2000s. It was argued
  that an initial state process was required along with complete
  congruence of steps (requiring total isolation at a high
  level)\cite{lisa0299}. The converse was argued: by creating closed
  operations in which the outcome was assured at a low level, the
  dependence on exterior ordering could be relaxed (which corresponds
  to a non-blocking execution policy)\cite{burgessC1,burgessC11}. The
  latter is just a micro-encapsulation of the former. The process is
  the same on different scales, but the latter is `reactive' in the
  sense of the Reactive Manifesto\cite{reactive}.}.  It's also the
approach used in pull requests, and GPS locators. The processes are
designed to favour a predetermined outcome. The outcome will only
become an event in the agent's future, and will only be observable as
a future event by other agents that depend on it.

On the interior of a process, a fixed point of a chain satisfies conditional promises:
\beq
A &\promise{+X_p | X_i}& A'\nonumber\\
A &\promise{+X_p | X_p}& A'.
\eeq

The more familiar retarded process is a Markov chain, to some order, and
has no deterministic end state unless the agents keep their promises
perfectly, which is essentially impossible to promise.

\subsection{Blocking and non-blocking promises}

Conditional promises are `blocking'. They are marked as `kept' only when a
precondition has been met. Unconditional promises are non-blocking.
In a sense, a promise of an advanced convergent state is a `blocking
algorithm'. The process exits when the final state has been reached.
In this case it is not waiting for input, as in blocking I/O, but
rather for the keeping of an interior promise. It doesn't refer to any
particular message, because it acts as a quasi-invariant condition. As
long as the condition is not met, nothing will proceed. If the process
drifts in and out of compliance, due to other subtime processes, then
blocking may add exterior latency.

\begin{figure}[ht] \begin{center}
\includegraphics[width=6.5cm]{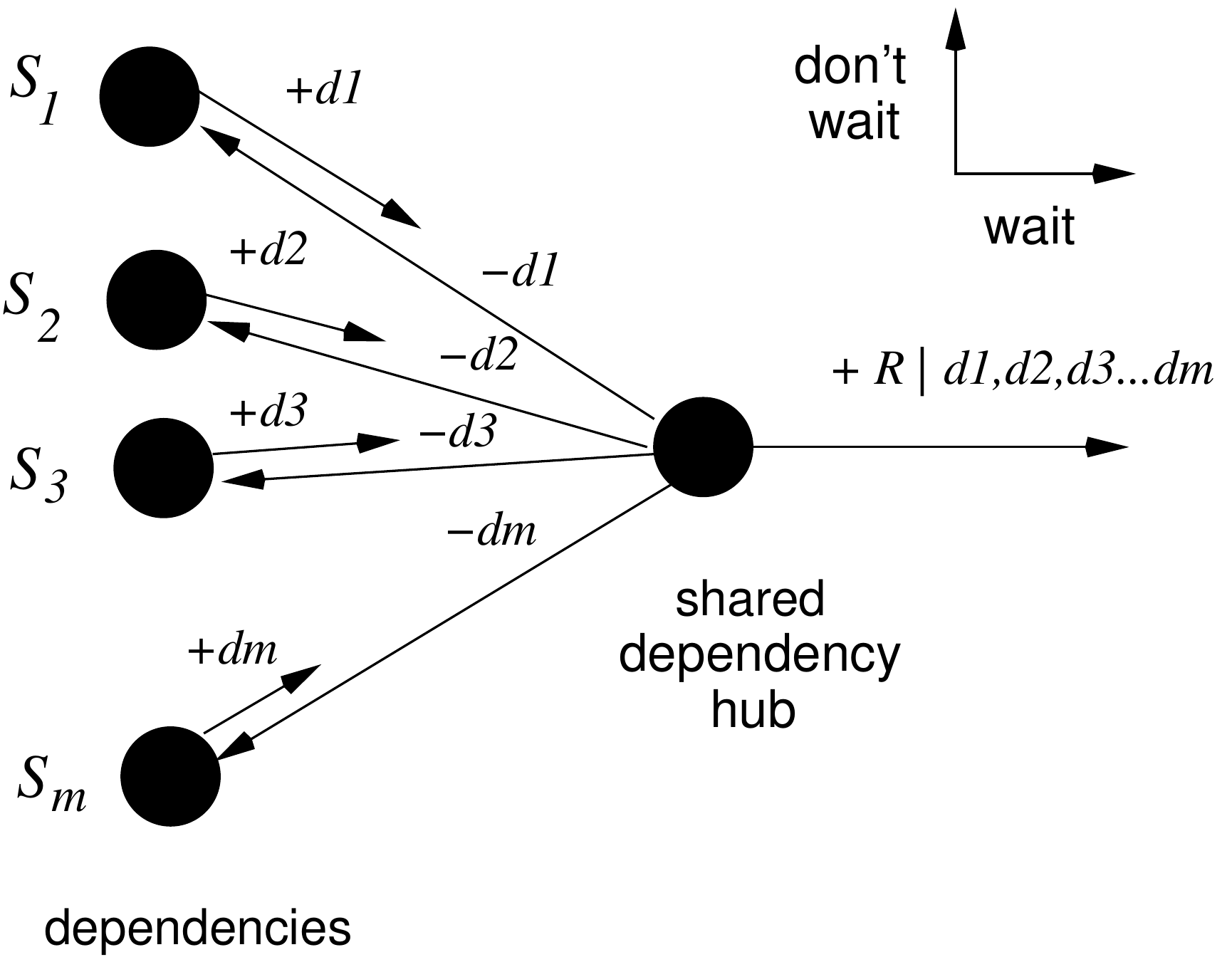} \caption{\small
Any active $N:M$ relationship will result
in partitioning of interior time at a hub. The granularity of that partitioning
is an interior policy decision for the agent. Coroutines, threads, and
serial blocking are three common variants of scheduling strategy for
sharing the agent's state machinery between the neighbouring
agents.\label{dependency2}} \end{center} \end{figure}

It only makes sense to speak of non-blocking in a shared time
environment, i.e. an agent that has interactions with more than one
dependency. Since each agent has its own process clock, the agent that
shares communications with these has to share its own clock with all
its dependencies---violating the `shared nothing' notion. Analogous to
reaching consensus equilibrium, any agent with multiple dependencies
does have to wait for all of them to keep their promises, else it
cannot keep its own promise. To summarize: an agent with multiple
dependencies need not wait for dependencies in any order (as they are
symmetrical with respect to the current promise, at scale $n$), but it
must wait for all of them in total, at scale $n+1$ (see figure
\ref{dependency2}).

I believe a more correct formulation of the Reactive Manifesto's call
for non-blocking processes\cite{reactive} is to decompose an agent
into subagents, i.e. partition them as non-ordered dependencies of larger
scale time-ordered dependencies  (see figure
\ref{dependency2}). In a single process, there must be a
hub that receives the results of such concurrent partitions
(`threads'), which must block on a larger scale to aggregate the
results.  A process (super)agent cannot be non-blocking on a scale of
exterior actions---at its exterior scale---without breaking such a
conditional promise. That would alter its causal behaviour. However,
the promise to depend on $m$ mutually independent dependencies should
not imply an arbitrarily imposed order; independent concurrent
processes need not be starved of a shared time resource because of the
need to wait for a subset of them.

\begin{example}
  A program need not suspend parallel threads or co-routines while one
  of them is waiting for a result. Waiting is a serial property, not a
  parallel one. The use of synchronous or asynchronous about
  communication implicitly uses the clock of one process to measure
  the progress of another, which has no invariant meaning.

  The reason for this nitpicking is that it may mislead readers into
  thinking that i) waiting is never necessary, and ii) all processes
  can be made more responsive by parallelization.  Not suspending
  parallel threads does not make a serial dependent thread
  asynchronous as measured according to its proper time. It may or may
  not be perceived that way according to some exterior clock time.
\end{example}
These points are certainly pedantic, but we need clarity as an
antidote to `best practices' that are merely advocated on the basis of
unclear language without explanation.

Convergence has no dependence on past state, as long as the final
outcome (desired end state) is a system invariant over each extended
epoch of the system.  It is an effective counter-strategy to the risk
of non-linear divergence.  For state that gets reused on multiple
occasions, it is a strategy for maintenance that builds {\em intrinsic
  stability} into systems\cite{treatise1}.

State convergence is also the way we render alternatives {\em
  indistinguishable} (by forgetting incidental past), and therefore
engineer greater stability through the fault tolerance. The instinct
to throw away the past may run counter to what many software developers are
trained to do---i.e.  to keep every distinct case separate in its own
context, but it actually leads to greater certainty in the future.
I'll go out on a limb and predict that we need a greater focus on
advanced causation for sustainable and scalable computing
in the future.

\subsection{Synchronous and asynchronous signals}

Synchronous means literally simultaneous---at the same time, i.e.
measured within the same clock tick interval---yet, when we speak of
synchronous or asynchronous communication, we are talking about serial
processes which (by definition) do not all happen at the same moment.
The only way to resolve this muddle is to define interior and exterior time
over coarse grained steps. This is effectively what we do in locking critical
sections in computer code. Many substeps of interior time can add up to a single step of
exterior time, which the exterior promise waits to be accepted and produce a tick.

In a distributed world of many clocks,
synchrony is a meaningless aspiration\cite{vectorclocks}. 
Synchronous can only mean `observed in the same interval', i.e. according to the same clock.
Time intervals are not invariant (they are covariant, i.e.  the change
with scale and observer reference frame).  Asynchronous implies that
an agent may wait for an unspecified interval after receiving a signal
before completing its dependent promise.

Both of these pertain to conditional promises:
\beq
A &\promise{Y}& S\label{start}\\
S &\promise{-Y}& A\label{precondition}\\
S &\promise{X | Y}& R\label{followup}\\
R &\promise{-X}& S\label{postcondition}\\
\eeq
Since we can only measure time differences locally at a single agent's clock,
a synchronous response would imply that the difference in $S$'s clock 
time between keeping promises (\ref{precondition}) and (\ref{followup})
was minimized.

An asynchronously-kept conditional promise would imply an arbitrary
delay between keeping promises (\ref{precondition}) and
(\ref{followup}). Thus synchronicity is a policy decision to set a
scale for a `timeout'. The semantics of faults also need to be
considered in these promises: a `fault' may also interpreted as a promise
outcome in this loose description.

The definition is more complicated when there is a shared channel (figure \ref{dependency2}).
\beq
S_1 &\promise{Y_1}& R\label{start2}\\
S_2 &\promise{Y_2}& R\\
\ldots\\
S_m &\promise{Y_m}& R\label{startm}\\
R &\promise{-Y}& S_1,S_2,\ldots,S_m\label{precondition2}\\
R &\promise{X | Y_1,Y_2,\ldots,Y_m}& A\label{followup2}\\
A &\promise{-X}& R\label{postcondition2}\\
\eeq
Event driven systems may be synchronous or asynchronous. This is a timescale issue.

\begin{lemma}[Synchronous or asynchronous events]
  \small The prerequisite for triggering a conditional promise is the
  sampling of all conditional events, no proper time interval for a
  response is implied by this order. 
\end{lemma}
A promise of a minimum response time after collection of dependencies
is limited by other (perhaps hidden) dependencies, e.g. CPU rate,
memory speed, scheduling commitments. It can be promised explicitly by $R$,
but has no absolute meaning for $A$, which samples according to its own clock.

Usually developers think about synchrony from their own perspective:
their viewpoint is that of an exterior observer (like a monitoring
system, not drawn in the figure), which has instantaneous knowledge of
the states of the agents. This is essentially a bad habit we take for
granted in everyday life, because we live in a relatively slow world in which signalling
is very fast.  The promise (\ref{start}) occurred when the final
result was `in the bank' $R$ according to its own clock.  This clock
is not usually distinguished from the clocks of the other agents, thus
effectively assuming a single global Newtonian view of time.  Alas, in
order to observe those agents directly, the same promise relationships
in figure (\ref{followup2}) are needed. Whether these are given
synchronously or asynchronously is scale dependent---a matter of
definition, not an observable fact, because it relies entirely on the
definitions of the final observer after it has sampled arriving
signals.  This is indeterminate, because to know this promise would be
to ignore the causal independence or violate the autonomy of the agents.

The implicit goal of the manifestos seems to be to render total systems as
close to causally deterministic as possible---recreating the Newtonian
view of global past. This is possible, but only at the expense of a
rate of total interior time the gets slower by at least $N^2$ for $N$
interior agents---as we know from consensus systems.

\section{Fault propagation}

Many developers believe that {\em modularity} prevents the propagation
of faults (hence the interest in microservices and object classes).
To isolate an agent from a dependency one must withdraw all promises
which use that dependency---abstaining might be an effective strategy
for the spread of consequences, but it also invalidates its purpose.
Locality may help to limit the propagation of a faults, if causes are
themselves modularized, and downstream clients make appropriate
promises to recover (see section \ref{downstreamsec}), but this is not
a certainty. It depends on how we define the semantics of separation.

The term `fault domain' is widely used to imply
a kind of semi-permeable membrane that prevents faults from having
consequences beyond a certain perimeter.  Security perimeters
(firewalls) are a common example. Such barriers may select only
specific messages from a wider set, but they cannot prevent the
propagation of influence unless there is independence of the promises
made by those modules.

A modular system may, on the other hand, help to pinpoint the source
of a fault, by attaching a name to a region, if the chain of causal
outcomes leave traces of the name in the states of agents as they propagate.

\subsection{Downstream principle}\label{downstreamsec}\label{cause}

Locality gives a surprisingly simple and consistent interpretation of
responsibility for keeping promises\cite{treatise2}. The recipient of
a promise carries the bulk of the burden of outcome.
The so-called {\em Downstream Principle}, in which agents have
responsibility for seeking alternatives when promised outcomes are not delivered
by upstream sources, follows from
the causal independence of agents, i.e. of agent autonomy.

In a chain of promises, dependencies are `upstream' (the servers or
sources of the flow of influence) and the benefactors or clients are
`downstream'. The assurance of the final promise outcome follows a
`downstream principle' that the agent farthest downstream has both
access and opportunity to observe and correct (or absorb) faults, and
hence the greatest causal responsibility for adapting to a promise not
being kept.  In other words, the greater the distance from the point
of promise-making, the less causal responsibility an agent has in
contributing to its outcome. Statelessness doesn't play a large role
in this principle; we only observe that the natural situation for
state is either far upstream or far downstream (at the ends of the
chain of dependency). This is tidy, for sure, so it helps developers,
but it also enables efficient scaling and a regularity of promised
patterns.

\begin{figure}[ht]
\begin{center}
\includegraphics[width=7cm]{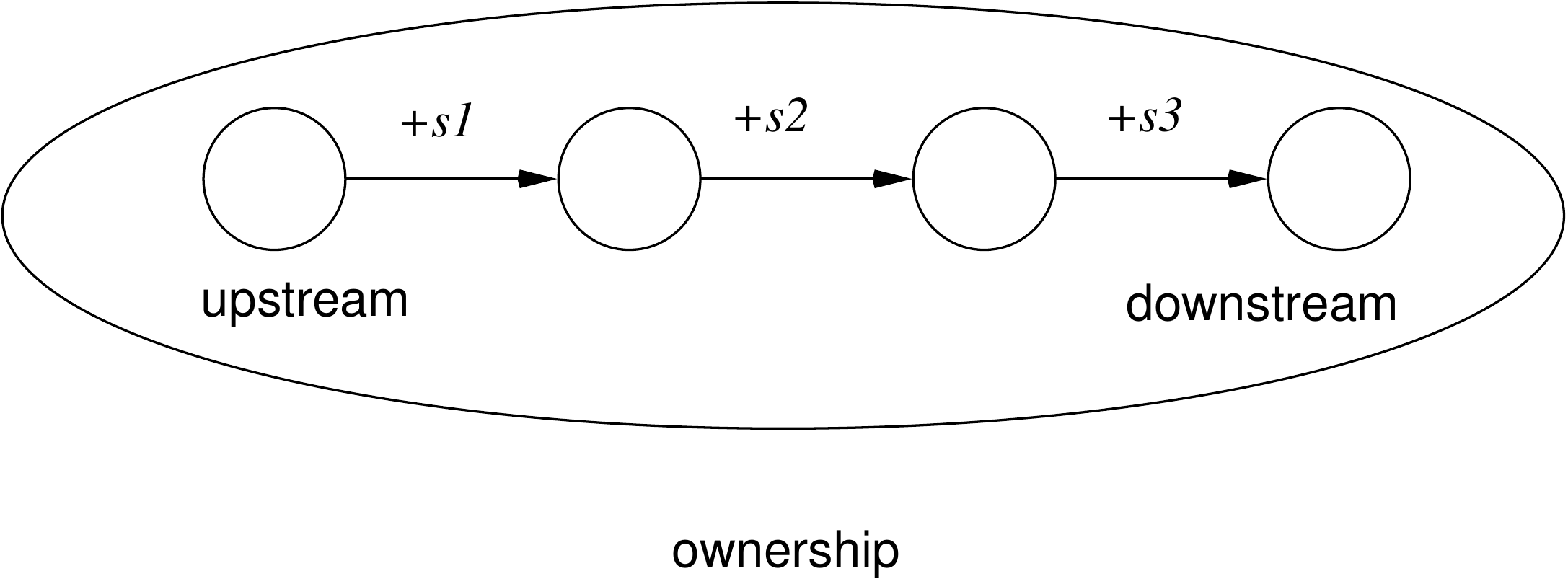}
\caption{\small Responsibility for success in a promise chain flows
  downstream. With feedback, upstream and downstream are
scale dependent concepts.\label{downstream}}
\end{center}
\end{figure}
This is not a moral assessment, it is a purely pragmatic observation
about cause and effect.  However, it is interesting that it is in
opposition to what is conventionally assumed about fault-tree
hierarchies and root cause analysis, which point a finger of blame at
the first choice of promise provider. The explanation for this
apparent contradiction can be found in the bi-directionality of
promise bindings required for propagation of influence.
By the conditional promise law, a promise that is conditional on
another promise being kept (either by the same agent or by a third
party) is not a promise, unless the other promise is made by the same
agent. This clarifies and documents diminished responsibility.

We can now attempt a limited but tenable definition of responsibility\cite{treatise2}:
\begin{definition}[Causal responsibility]
\small
  When an agent relies on a dependency promise in order to keep its
  own conditional promise, causal responsibility refers to the agent's
  freedom to obtain a promised outcome by its own autonomous choice of
  interaction, especially in the presence of redundant alternatives.
\end{definition}

In Promise Theory, we track provenance, or causation with {\em
  conditional promises}, as chains of promises.  Keeping each promise is the
responsibility of the agent that makes the promise (the promiser).
However, from the conditional promise law, an agent making a conditional
promise has not made a complete promise at all unless it also promises to
acquire the thing its promise is conditioned on.
\begin{figure}[ht]
\begin{center}
\includegraphics[width=6cm]{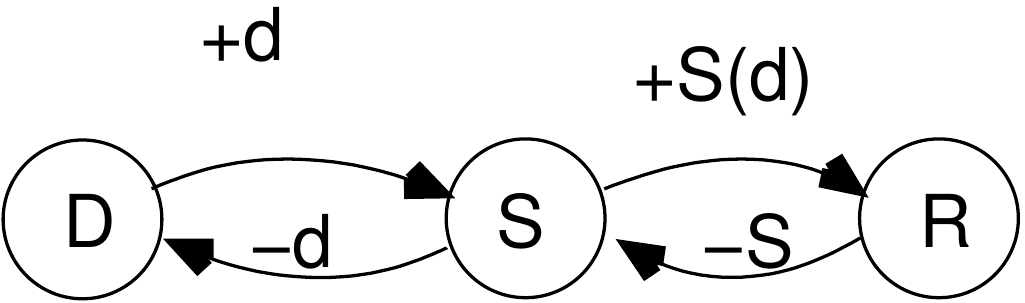}
\caption{\small A conditional promise chain, showing service delivery based on
a dependency.\label{cond1}}
\end{center}
\end{figure}
Thus the promise depends on the promises of other agents---a shared responsibility,
which makes the promise more fragile, as it depends on the promises of both the first
agent AND the second.
Consider the scenario in figure \ref{cond1}
\beq
D &\promise{+d}& S\\
S &\promise{-d}& D\\
S &\promise{+S(d)}& R\\
R &\promise{-d}& S
\eeq
where
\beq
S \promise{+S(d)} R \equiv
\left\{
\begin{array}{c}
S \promise{+S|d} R\\
S \promise{-d} R
\end{array}
\right.  \eeq This system is fragile because the recipient has only a
single choice. It has a single point of failure. If The recipient
could seek out redundant alternatives to provide the service $S$.
Nothing can improve the situation for these agents from outside, since
what happens beyond the horizon of the next agent in the chain of
promise relationships is beyond the control of the recipient, and is
thus beyond the limit any possible responsibility, but there is the
possibility to improve for the client by promising redundancy on a
larger scale.

\begin{figure}[ht]
\begin{center}
\includegraphics[width=6cm]{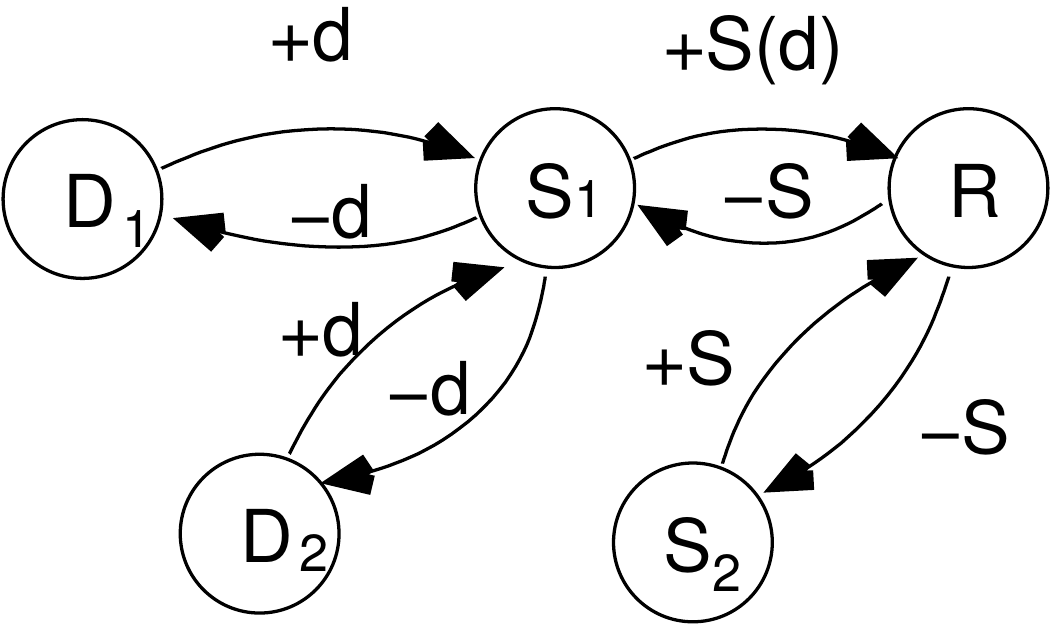}
\caption{A redundant conditional promise chain, showing service delivery based on
a dependency.\label{cond3}}
\end{center}
\end{figure}

Now consider the same scenario with redundancy along the chain (see
figure \ref{cond3}) \beq
D_1 &\promise{+d}& S_1\\
D_2 &\promise{+d}& S_1\\
S_1 &\promise{-d}& \{D_1,D_2\}\\
S_1 &\promise{+S|d}& R\\
S_1 &\promise{-d}& R\\
S_2 &\promise{+S}& R\\
R &\promise{-S(d)}& S_1\\
R &\promise{-S}& S_2\\
R &\promise{-d}& \{S_1,S_2\}
\eeq 
In this second scenario, both the
server $S_1$ the recipient can choose from two providers of the
promises they are trying to use.  For the final recipient $R$, the
fact that the promise from $S_1$ has a dependency is irrelevant, as
there is nothing it can do about that except to acquire a second
provider who may or may not have a dependency too.
The only security the recipient $R$ has is to have a choice of
providers. No matter how hard the providers $S_1$ and $S_2$ try to
keep their promises of service, unforeseen circumstances may prevent
them from doing so. Indeed $R$ may itself be negligent receiving their
services.

This suggests that, while responsibility for keeping a promise lies with
each source agent, only the final recipient can be considered
responsible for securing a successful promise outcome. It's up to the
client to acquire alternative sources, so any state should commute
across these parallel alternatives. This can be handled by avoiding
state, by leaving state with the client, or by arranging for {\em
  consensus} about state\cite{paxos,raft}.

We now see the concision and utility of the promise theoretic view. 
The assisted promise law states that \cite{promisebook}: if an
interior promise $\pi$ by an agent $A$ is converted into an assisted
promise $\pi'$, through an intermediate agent $I$, we can draw a new
boundary around $\{ A,I\}$ and no memoryless process will be able to
distinguish them, since the semantics of the promises are equal and
any difference in the dynamical response times of the promise keeping
is not observable without reference measurements to calibrate with.

\section{Applications}

Given the length of the paper already, let's only briefly try to
review the application of the principles described in the paper to
some key IT areas. Software systems are a superposition of several
hierarchically dependent processes that may change independently:
\begin{itemize}
\item Code developments change.
\item Version deployment changes.
\item Operational runtime changes.
\end{itemize}
The goal of a developer or system architect is to list the
invariant and variant characteristics of a system over
its separable timescales.

\subsection{Data pipelines in the extended cloud (IoT)}

The extended cloud consists of datacentre services and the coming edge
services, where data are collected, known as the Internet of Things
(IoT).  It comprises fast localized resources where users compete for
shared services (Amazon, Google, Microsoft Azure, etc), and slower
delocalized resources which are naturally partitioned and where data
originate (mobile phones, home computing, distributed sensor nets,
etc).

The challenges involve processing data where resources are available,
avoiding the movement of large amounts of data unnecessarily.
\begin{itemize}
\item Invariant policy for aggregation when data come together at a hub.
\item Invariant promises between independent stages of the pipeline.
\item Invariant handling of message promises with non-deterministic arrival.
\item More.
\end{itemize}

\subsection{Microservices}

The development architecture known as microservices has become popular
for cloud native development\cite{microservices}, contrasted with the
perjorative `monolith'. The name seems to refer to the separation of
modules as formally independent processes that may be hosted by
different cloud instances. The key innovation here is in sharding the
development process (not just the code) to enable development agility
and full lifecycle ownership of code\footnote{Previously it was common
  for development and operations to be dealt with by separate teams
  with weak ties. Microservices embodies an implicit promise for both
  phases to be managed by the same team, or equivalently by different
  members with strong ties.}. It should be clear that this is a non-invariant
characterization of a system, which depends on fixing a particular scale.
If we increase the boundary of a set of modules to incorporate all their
interactions, we get a new monolith on a larger spatial scale, and on a longer
timescale.

Applying the principle of separation of timescales for approximate invariance of promises.
\begin{itemize}
\item All developer team processes tick on different clocks (version
  control submissions may be counted as ticks of the development
  clock, for instance). Every hub that integrates changes combines
  clocks into policy based outcome.

\item Code is invariant when it refers to identical source code: changes to
any part of the code, including all dependencies, as well as platform
dependencies, count as a new version. A proper set of promises would include
a packaging of all dependencies. Due to the separation of containers and
platform this may not be practical today. With future unikernel platforms,
this would be realizable down to the level of the processor hardware.

\item A code base can vary independently of the platform in which it runs.
Similarly, the platform may vary independently of the codebase as a
dependency.

\item Latency between communicating processes applies to the runtime communication and to
code changes, where dependencies are involved. If the promises made by one service change,
downstream processes also have to adapt.

\end{itemize}

Modular separation of service instances reduces dependencies between
interior details of team-work, and refactors or renormalizes it into computer code
dependencies.  Key promises are moved from exterior to interior of
agents. Whether promises concern the interior or exterior of a certain
superagent seems not to matter to the outcome of the arbitrary
boundary of the system, since the boundary choice is not an invariant.
It might matter to the teams and the runtime efficiency of course.

\subsection{Manifesto promises}

Some brief comments on some of the manifesto promises that developers
are encouraged to keep:
\begin{itemize}
\item Elasticity is the promise to spawn new agents and feed them data
  at a rate determined by the width of the bottleneck.

\item Responsiveness is a promise to continue in a `timely manner', i.e.  to
  minimize the response time for imposed requests. To keep the promise
  of a consistent response time assumes that the response is invariant
  under changes in the message size, etc. When agents collectively
  respond to messages, there may need to be coordination and scaling
  of the messages to inform every partial agent of an arrival
  (so-called domain events).

\item Message based design does not make a system responsive, because
  the sending of a message is not a driver of the response, the
  sampling of the message is.  Event driven systems may be synchronous
  or asynchronous, they are just conditional promises.  What makes a
  system responsive is the policy (promise) to schedule a dependent
  agent `quickly' on receiving a message. The challenge is to define
  whose clock gets to decide what `quickly' means.

\item If a promised output depends on receiving multiple messages,
  from the same or different sources (as is common in data pipelines),
  then the policy needs to include specifics about how messages
  arriving at possibly different rates will be combined and ordered
  (see the Koalja pipeline, for example\cite{koalja}).

\item Asynchronous messaging is a policy choice, which informally
  implies a buffer queue between agents, decoupling them at the source
  rather than imposing on them at the receiver end. A strict
  synchronous system was have to drop many messages because no two
  systems can be fully synchronous unless downstream processing
  bandwidth is always greater than upstream, at every stage (like a
  river delta).

\item Failures are contained within each component, isolating components
from each other and thereby ensuring that parts of the system can fail
and recover without compromising the system as a whole. Recovery of
each component is delegated to another (external) component and
high-availability is ensured by replication where necessary. The
client of a component is not burdened with handling its failures.

\item Reactive Systems\cite{reactive} can adapt to changes in the
  input rate by increasing or decreasing the resources allocated to
  service these inputs. This implies designs that try to eliminate contention
  points or central bottlenecks, resulting in the ability to shard or
  replicate components and distribute inputs among them. Reactive
  Systems support predictive, as well as Reactive, scaling algorithms
  by providing relevant live performance measures.

\item Reactive Systems rely on asynchronous message-passing to establish a
boundary between components that ensures loose coupling, isolation and
location transparency. This boundary also provides the means to
delegate failures as messages. Employing explicit message-passing
enables load management, elasticity, and flow control by shaping and
monitoring the message queues in the system and applying back-pressure
when necessary. Location transparent messaging as a means of
communication makes it possible for the management of failure to work
with the same constructs and semantics across a cluster or within a
single host. 

\item Redundancy enables a client for whom a transaction fails to
  promise a `retry' from its side of a binding, with an alternative
  provider. The redundant choices only need to be `good enough
  alternatives' to satisfy a client.  In practice, they might be
  subtly different versions, a different model of vehicle (`I'm sorry
  we don't have your first choice today').  In reliable services, like
  TCP, users are guaranteed some kind of a response (but not a
  delivery time), but and the stateful nature of the delivery
  mechanism (which routes are taken by packets) are not revealed to
  the end users. That doesn't make TCP stateless, but it just shifts
  the responsibility for keeping state into a shared responsibility,
  in which the client assumes its natural downstream role.

\item Consistency. 
In the IT industry, the responsibility for promise outcomes is almost uniquely apportioned
to the service provider---we require the promise of data
consensus between alternative providers, which is impossible in
general over a finite interval of arbitrary time. A safer strategy is
to render clients insensitive to the variations amongst redundant
alternatives, using intrinsic stability of fixed points, and
idempotent operations (see section \ref{idemp}). One tries to make
redundant agents indistinguishable from one another
\cite{observability}. One way to do this is to remove their dependence
on interior state. However, purely stateless behaviour is neither
efficient nor desirable because it doesn't take care of adapting to
user needs without multiplying every possible combination of choices
as statically independent pathways through a system. This applies to
the arguments about `immutability' too. The cost of redefining state
from being runtime to an initial condition may be high.

\end{itemize}

\section{Summary}

The reproducibility and functional stability of systems of agents
depend on a few key principles, of which the {\em separation of
  dynamical scales} is the most important.  In systems engineering,
the engineer basically figures out how to distribute a collection of
process promises using the criteria of state localization and
longevity alongside a graph of conditional causal influence.
Popular discourse is imprecise in its terminology. This paper
offers a set of concepts that are not wrong, which could be
used as a reference model.

On the matter of statelessness, non trivial processes can never be
fully stateless, but independence of certain states over certain
regions of a system can be strategically motivated.  When we talk
about statelessness, there is an implicit downstream observer in the picture---an
agent that will receive an outcome that acts in the role of
observer---perhaps a service client. When a service is stateless, the
client cannot distinguish outcomes based on its history of prior
interactions (the client doesn't make a mark or leave a dent in the
server!).  The same principle applies client-server interactions like
classic Web applications, and to data pipelines\cite{koalja}. In
either case one is definitely interested in being able to store data
with integrity in a service: storage servers, filesystems, databases,
and caches all make this promise explicitly.  The goal is to enable
stateful behaviour without loss of efficiency, continuity, or
stability on the scale of a total application. By making use of agent
autonomy, developers often try to isolate what are commonly referred
to as `fault domains'.

From what I can tell, statelessness is something of a red herring in
the story of reliability and scalability, as it's not an invariant
characterization; it refers to a preferred scale and viewpoint.  The
proper invariants in a process are the promises, including conditional
switching rules, that link process agents into a system, and how
interactions are localized during their keeping.  Rhetoric aside, the
actual goal of `statelessness' seems not to be to abhor state, but
rather to {\em contain} it inside transactional elements, whose
outcome is not trusted until the promised outcome has been assessed as
kept. Containment has nothing to do with fault localization; rather, 
it is a prerequisite for transactional change to be preserved together
for long enough to complete the transactions under invariant
conditions.

Localization of state (the scope of memory behaviour) concerns a tradeoff
between decision flow and resources:
\begin{itemize}
\item {\em Locally stateless}: means `memoryless', which may imply
  limited transactional invertibility, but long-lived stability and
  sustainability. We would be better served by noting which agents
are memoryless in their transformations.

\item {\em Stateful}: means `memory process'. 
Which changes depend not only on incoming data but on the current
state of the agent?
All processes are
  stateful on some scale. The stateful parts bring potential for fragility and
  possibly non-linear behaviour, and may be unsustainable unless there
  is voluntary obsolescence of history\footnote{While memory is crucial
  for some tasks, it shouldn't be accumulated without good
  reason---that's called a memory leak.}.

\item {\em Fixed point behaviour}: a memory process referring to
an invariant future state, with intrinsic process stability,
  which does not require any runtime memory, a priori, only a
  maintenance process that counters state drift and converges, or the
  absence of complete isolation from external change\cite{burgesstheory,burgessC11}.

\end{itemize}
The strategy of engineering around fixed points still goes highly
unappreciated across software engineering and
management\cite{treatise1}.  The legacy of industrial commoditization
is still with us, and out old-fashioned thinking favours the pattern
of replacing defective parts with fresh `clean' parts (like changing
the air filters). This process favours the builder, but may be a
wasteful strategy for the system as a whole, especially when repair
can be automated cheaply.  Because state drift can't be prevented in
practice, we need either maintenance over some timescale, or voluntary
obsolescence (apoptosis) of process.  There will need to be a greater
focus on the benefits of advanced causation to scale sustainable in
the future.  Disposability of systems and their runtime state in
processes is linked to transactional breakdown of process, which in
turn is linked to a message strategy. There is nothing wrong with
these approaches, but they may not be substantially better from all
viewpoints---they favour a developer viewpoint rather than a client
viewpoint, a system viewpoint, or a sustainability viewpoint.

The remaining freedoms in process design lie in the avoidance of
serial contention (as Amdahl's law) and mutual coherence (distributed
consistency), by partitioning activity into independent timelines.
Locality of dynamics through 
{\em service access points} addresses the trade-off between space and time:
\begin{itemize}
\item {\em Parallel or partitioned agents}: implies shorter queueing, per agent
  or partition, at the expense of more agents over more space to configure,
  maintain, and power.

\item {\em Serialised monolithic agents}: implies fewer agents,
  i.e. less space used, perhaps at the expense of longer response time due to
  contention in queues or collisions.
\end{itemize}
This is not the same as the decomposition of semantics in code within
boundaries.

Finally, let me mention a few words about memory processes 
given the current focus on {\em learning systems}, such as 
in so-called `Artificial Intelligence', etc. Any learning system is a memory
process, by design. This also applies to any system
that keeps memory that may exert an influence over causally related
outcomes.  Reasoning processes are state machines, by any
measure---attempting to describe them as purely ballistic
transactional phenomena, by focusing on only a small part of their
processes, only delays inevitable consequences.

Locality of promises, at a stated scale, may be the preferred way to
describe behaviours relative to the cost and availability of
collaborative resources.  The unspoken assumption in a lot of cases is
that `local interior resources' are cheaper and faster than `non-local
remote resources', and that `remote resources' are safer from failures
by spreading the potential for failures over a wider area. This has
all been turned upside down several times in the virtualized world
though. New technologies alter these basic assumptions frequently, so
we need an approach based on the idea that the perturbations leading
to faults are themselves phenomena with scales of their own.

{\bf Acknowledgment:} I'm grateful to Kenny Bastani, Jonas Bon\'er, Andrew Cowie,
and Daniel North for commenting on drafts and offering
references. Thanks also to Colin Breck and Tristran Slominski for a thorough reading.

\appendix

We can supply an invariant definition of memorylessness using
stochastic processes.  Statelessness refers to memorylessness in the
sense of a Markov process: the dependence on current state is
inevitable as long as there is input to a process, but dependence of
behaviour on an accumulation of state over many iterations is what
people usually mean by stateful behaviour.

Variables are embedded agents that keep simple promises to remember a value.
A stateful process accumulates parametric data over a number of
interior times $t_0, t_1, t_2, \ldots$, so that we could write
each promise made by the process as a function of all those times:
\beq
\pi&:& A \promise{+V[d(t_0), d(t_1), \ldots]\; |\; d(t_0), d(t_1), \ldots} A'\nonumber\\
&\simeq& A \promise{+V[d(\overline t)] \;|\; d(\overline t)} A' \eeq which schematically means
that \beq
\frac{\partial \pi}{\partial d} &\not=&0,\\
\frac{\partial \pi}{\partial \overline t} &\not=&0.  \eeq i.e. the promise made
by the agent is not constant over the interactions it promises. The
memory $d({\overline t})$ is accumulated from some initial time, making the
promise evolve. This is called a {\em memory process}. The converse of
a memory process is a memoryless process. A memoryless process may be
constant in time, or it can depend on only that last known state, like
a ballistic trajectory (imagine billiard balls whose change in
behaviour is entirely determined by the last ball to strike
them)\footnote{This is an interesting example because in Newtonian
  mechanics, a collision may be memoryless, but the trajectory isn't.
  The momentum of balls is conserved and remembers the sum effect of
  prior collisions. This is one of many ways to illustrate how memory
and state are scale dependent.}.

Memoryless processes are also called Markov processes (see figure \ref{scattering}).
Their behaviour is `ballistic' in the sense that the arrival of a prerequisite state
effectively triggers the release of what is promised by each agent.
A Markov process is usually described as a chain of agents that
satisfy a condition about random processes, and based in probabilities\cite{grimmett1}.
\begin{definition}[Markov Chain]
\small
Let $X_n$ be a discrete random variable, for non-negative integer
$n=0,1,2,\ldots$, taking values in $\{x\}$.
\beq
P(X_n = s\; |\; X_0 = x_0,  \ldots, X_{n-1} = x_{n-1}) =\nonumber\\
  P(X_n = s \;|\; X_{n-1} = x_{n-1}) ~~~~~~ ~~~~~~~
\eeq
for all $n \ge 1$ and $s \in \{x\}$.
A Markov process has a {\em transition matrix} or
{\em scattering matrix} for the discrete set of states $X_n$:
\beq
T_{ij}^{(n)} = P(X_{n+1}=j | X_n=i)
\eeq
If this transition matrix is independent of $n$, i.e. $T_{ij}^{(n)}=p_{ij}$, for all $n$,
then the chain is said to be homogeneous or translationally invariant.
\end{definition}
Probabilities, in the usual sense are globally defined, but we can
replace them with {\em assessments} in Promise Theory, which are the local
equivalent. Each observer agent $O$ in a system may form its own
assessment $\alpha_O(\pi)$ of the probability that a promise $\pi$
will be kept, for any definition of probability. Then the above
definitions apply for any local observer by the association: 
\beq
O_{ij}^{(n)} = \alpha_O(X_{n+1}=j | X_n=i).\label{markov}
\eeq

In a Markov scattering process, each input leads to a unique output by
a fixed rule. The scattering doesn't depend on the order of the inputs
nor their relative frequencies. The scattering matrix does not remember
past inputs; everything depends on the last one. This makes the scattering
agent autonomous or causally independent\footnote{This may also be used as
a definition of the autonomy of the agents.}.
\begin{definition}[Causally independent]
\small An agent is causally invariant under a promised influence $x$ if
it does not depend on a parameter $x$, so that.
\beq
x \rightarrow x' \therefore V(x) = V(x'), ~~~\forall x,x'.
\eeq
\end{definition}
In a quasi-differential shorthand, we might also be tempted to write:
\beq
\frac{\partial V}{\partial x} = 0.
\eeq
Finally, we should note that this should not be taken to mean that $V$ is differentiable, as no
such mathematical property exists in the real world, but we can
construct state space extended generalizations that include averaging,
and so on, so I'll ask the forbearance of readers and follow common
practice and use this as a shorthand for the expression of
independence of $V$ on some parameter $x$. For a popular discussion of
the meaning of this, see \cite{certainty}.

\bibliographystyle{unsrt}
\bibliography{spacetime}

\end{document}